%% file: paper.tex
\documentclass[prd,twocolumn,showpacs,showkeys,preprintnumbers,floatfix,
  nofootinbib,superscriptaddress]{revtex4-2}
\usepackage{amsfonts} % AMS
\usepackage{amssymb} % AMS
\usepackage{amsmath} % AMS
\usepackage{mathrsfs} % AMS
\usepackage{subfigure}
\usepackage{graphicx} % Include figure files
\usepackage{array} % array
\usepackage{dcolumn} % Align table columns on decimal point
\usepackage{bm} % bold math
\usepackage{latexsym} % latex symbols
\usepackage{longtable} % long tables
\usepackage{multirow} % multiple rows merged to a row
\usepackage{hyperref} % hypertext links 
\usepackage{textcomp}
\usepackage[usenames,dvipsnames,x11names]{xcolor}
\usepackage{mathrsfs}
\usepackage{comment}
\usepackage{rotating}
\usepackage{here}
\usepackage{float}
\usepackage{multirow}

\input{macro.tex}

\graphicspath{{./figs/}}

\allowdisplaybreaks

\begin{document}

\title{ 2025 update on $\varepsilon_K$ in the Standard Model
  with lattice QCD inputs}

%\author{Jon A. Bailey}
%
%\affiliation{
%  ISED, UIC, Yonsei University,
%  Incheon 21983, South Korea
%}

%\author{Yong-Chull Jang}
%
%\affiliation{
%  Physics Department,
%  Brookhaven National Lab, Upton,
%  NY 11973, USA
%}

\author{Seungyeob Jwa}
\affiliation{
  Lattice Gauge Theory Research Center, FPRD, and CTP, 
  Department of Physics and Astronomy,
  Seoul National University, Seoul 08826, South Korea
}

\author{Jeehun Kim}
\affiliation{
  Lattice Gauge Theory Research Center, FPRD, and CTP, 
  Department of Physics and Astronomy,
  Seoul National University, Seoul 08826, South Korea
}

\author{Sunghee Kim}
\affiliation{
  Lattice Gauge Theory Research Center, FPRD, and CTP, 
  Department of Physics and Astronomy,
  Seoul National University, Seoul 08826, South Korea
}

\author{Sunkyu Lee}
\affiliation{ Center for Precision Neutrino Research (CPNR),
  Department of Physics, Chonnam National
  University, Yongbong-ro 77, Puk-gu, Gwangju 61186, Korea
}

\author{Weonjong Lee}
\email[E-mail: ]{wlee@snu.ac.kr}
\affiliation{
  Lattice Gauge Theory Research Center, FPRD, and CTP, 
  Department of Physics and Astronomy,
  Seoul National University, Seoul 08826, South Korea
}

%\author{Jaehoon Leem}
%
%\affiliation{
%  Computational Science and Engineering Team, Innovation
%  Center, Samsung Electronics, Hwaseong, Gyeonggi-do 18448, South
%  Korea
%}

\author{Sungwoo Park}
\affiliation{
  Lawrence Livermore National Lab,
  7000 East Ave, Livermore, CA 94550, USA
}

\collaboration{SWME Collaboration}

\date{\today}

\begin{abstract}
%
\begin{comment}
We present theoretical results for the indirect CP violation
parameter, $|\epsK|$ calculated directly from the standard model using
lattice QCD inputs such as $\BK$, $\Vcb$, $\Vus$, $\Vud$, $\xi_0$,
$\xi_2$, $F_K$, and $m_c$ (charm quark mass).
%
We find a strong tension in $|\epsK|$ at the $\approx 5\sigma$
($5.2\sigma \sim 4.6\sigma$) level between the experimental value and
the theoretical value calculated directly from the standard model
using lattice QCD inputs.
%
The standard model with lattice QCD inputs describes only 65\%
of the experimental value of $|\epsK|$, and does not explain its
remaining 35\%.
%
We also find that this tension disappears when we use inclusive $\Vcb$
which comes from the heavy quark expansion and QCD sum rules.
%
This tension is highly correlated with the discrepancy between
exclusive $\Vcb$, and inclusive $\Vcb$.
%
We also present results for $|\epsK|$ obtained using the
Brod-Gorbahn-Stamou (BGS) method of $u-t$ unitarity, which leads to
even a stronger tension at the $5.5\sigma \sim 4.9\sigma$ level with
lattice QCD inputs.
\end{comment}
%
We present theoretical results for the indirect CP violation
parameter, \( |\varepsilon_K| \), evaluated directly within the
Standard Model using lattice QCD inputs including \( B_K \), \(
|V_{cb}| \), \( |V_{us}| \), \( |V_{ud}| \), \( \xi_0 \), \( \xi_2 \),
\( F_K \), and the charm quark mass \( m_c \).
Our analysis reveals a significant tension—at the \( \sim 5\sigma \)
level (\( 4.6\sigma \) to \( 5.2\sigma \))—between the Standard Model
prediction and the experimental value of \( |\varepsilon_K| \).
The Standard Model prediction, informed by lattice QCD inputs,
accounts for only approximately 65\% of the experimental value,
leaving a 35\% discrepancy unexplained.
Notably, this tension vanishes when using the inclusive determination
of \( |V_{cb}| \), derived from heavy quark expansion.
The discrepancy is therefore tied to the well-known tension between
the exclusive and inclusive determinations of \( |V_{cb}| \).
We further report results for \( |\varepsilon_K| \) obtained using the
Brod--Gorbahn--Stamou (BGS) method based on \( u\text{--}t \)
unitarity, which yields an even stronger discrepancy, in the range of
\( 4.9\sigma \) to \( 5.5\sigma \), when combined with lattice QCD
inputs.
\end{abstract}

\maketitle

\input{intro.tex}

\input{master.tex}

\input{input.tex}

\input{result.tex}

\input{conclude.tex}

\begin{acknowledgments}
  We would like to express our sincere gratitude to Andreas Kronfeld,
  Aida El-Khadra, Maarten Golterman, and Stephen Sharpe for helpful
  discussion.
  We also would like to express sincere gratitude to Guido Martinelli
  for providing to us the most updated results of UTfit.
  The research of W. Lee is supported by the Mid-Career Research Program
  (Grant No. NRF-2019R1A2C2085685) of the NRF grant funded by the
  Korean government (MSIT).
  W. Lee would like to acknowledge the support from the KISTI
  supercomputing center through the strategic support program for the
  supercomputing application research (No. KSC-2018-CHA-0043, KSC-
  2020-CHA-0001, KSC-2023-CHA-0010, KSC-2024-CHA-0002).
  Computations were carried out in part on the DAVID cluster at Seoul
  National University.
\end{acknowledgments}

%-----------
% reference
%-----------
\bibliography{ref}
\end{document}

%% file: macro.tex
\providecommand{\wbar}[1]{\overline#1}

\providecommand{\mate}[3]{\langle#1\lvert#2\rvert#3\rangle}

\renewcommand{\Re}{\mathrm{Re}\,}
\renewcommand{\Im}{\mathrm{Im}\,}

\providecommand{\GeV}{\mathrm{GeV}}
\providecommand{\MeV}{\mathrm{MeV}}

\providecommand{\CL}{\nonumber\\}

\definecolor{HLBlue}{HTML}{6599FF}
\definecolor{HLOrange}{HTML}{FF6600}

\newcommand{\MSb}{\overline{\mathrm{MS}}}
\newcommand{\BGS}{{\mathrm{BGS}}}

\providecommand{\CL}{\nonumber\\}
\newcommand{\cB}{\mathcal{B}}

\newcommand{\CS}{\mathscr{S}}

\newcommand{\CKM}[1]{|V_{#1}|}

\newcommand{\Vcb}{\CKM{cb}}
\newcommand{\Vub}{\CKM{ub}}
\newcommand{\Vus}{\CKM{us}}
\newcommand{\Vud}{\CKM{ud}}

\newcommand{\BK}{\hat{B}_{K}}

\newcommand{\eps}{\varepsilon}

\newcommand{\epsK}{\varepsilon_{K}}

\newcommand{\BtoDst}{\bar{B} \to D^\ast \ell \bar{\nu}}
\newcommand{\BtoD}{\bar{B} \to D \ell \bar{\nu}}

\newcommand{\RBF}{R_\text{BF}}
\newcommand{\Lbtop}{\Lambda_b \to p \; \ell^- \bar{\nu} }
\newcommand{\LbtoLc}{\Lambda_b \to \Lambda_c \ell^- \bar{\nu} }
\newcommand{\BstoK}{B_s \to K^- \ell^+ \nu }
\newcommand{\BstoDs}{B_s \to D_s^- \ell^+ \nu }
\newcommand{\BstoDsst}{B_s \to D_s^{\ast -} \ell^+ \nu }

\newcommand{\red}[1]{\textcolor{red}{#1}} 
\newcommand{\blue}[1]{\textcolor{blue}{#1}} 
 
\newcommand{\greenfour}[1]{\textcolor{Green4}{#1}} 
 
\newcommand{\magenta}[1]{\textcolor{magenta}{#1}}

\newcommand{\black}[1]{#1} 

%% file: intro.tex
\section{Introduction}
\label{sec:intr}
%------------------
% opening sentence
%------------------
CP violation offers a uniquely clean and powerful probe of physics
beyond the standard model (SM) \cite{ Buchalla:1995vs, Buras:1998raa}.
In the neutral kaon system, CP violation manifests through two
distinct mechanisms: direct CP violation and indirect CP
violation.\footnote{ Good reviews on this subject might be found in
Refs.~\cite{ branco1999cp, Sozzi:2008zza, Buras:2020xsm,
  Buchalla:1995vs, Buras:1998raa, Lellouch:2011qw}. }
The focus of this work is on the latter, which is encapsulated by
the parameter $\epsK$, defined as 
\begin{equation}
  \label{eq:epsK_def}
  \epsK 
  \equiv \frac{\mathcal{A}(K_L \to \pi\pi(I=0))} 
              {\mathcal{A}(K_S \to \pi\pi(I=0))} \,,
\end{equation}
where $K_L$ and $K_S$ are the long- and short-lived neutral kaon
states, and $I=0$ denotes the isospin of the final two-pion state.
Experimentally, the value of $\epsK$ is measured to be \cite{
  ParticleDataGroup:2024cfk},
\begin{align}
  \label{eq:epsK_exp}
  \epsK &= \left[ (2.228 \pm 0.011) \times 10^{-3} \right] 
  \times e^{i\phi_\eps} \,,\CL
  \phi_\eps &= 43.52 \pm 0.05 {}^\circ \,.
\end{align}
indicating that the $K_L$ state contains a small ($\approx 0.2\%$)
admixture of CP-even components, with the remainder being dominantly
CP-odd.

Among the CP-violating observables in kaon physics, $\epsK$ stands out
as an especially sensitive probe for new physics, for two key
reasons.
First, its experimental determination is extremely precise
\cite{ ParticleDataGroup:2024cfk}.
Second, advances in lattice QCD now allow theoretical predictions with
commensurate precision, enabling stringent comparisons between theory
and experiment \cite{Meinel:2024pip}.
%

%------------
% why e_K?
%------------
In the SM, CP violation arises from a single complex phase in the CKM
matrix.
Thus, any additional sources of CP violation beyond the SM can
lead to observable deviations in $\epsK$.
Consequently, high-precision determinations of $\epsK$ within the SM
framework—particularly using lattice QCD inputs—have become a top
priority in the field \cite{Lee:2006cm, Feng:2017voh}.

%-----------------------
% brief history on e_K
%-----------------------
A complete SM prediction of $\epsK$ requires knowledge of
approximately 18–20 input parameters \cite{Bailey:2015tba}, of which
nine can, in principle, be determined using lattice QCD: $\BK$,
$\Vcb$, $\xi_0$, $\xi_2$, $\xi_\text{LD}$, $\Vus$, $\Vud$, $m_c(m_c)$,
and $F_K$.\footnote{ Here $\xi_0$ and $\xi_2$ are redundant.  We need
to know only $\xi_0$, but it is possible to obtain $\xi_0$ from
$\xi_2$ using $\eps'/\eps$. For more details, refer to Subsection
\ref{ssec:xi0}. }\footnote{ Here $\Vus$ and $\Vud$ are redundant. We
need to know only the ratio $r = \Vus/\Vud$. For more details, refer
to Subsection \ref{ssec:wp}.}
Among these, $\BK$ is now known with remarkable precision (1.2\%
uncertainty) thanks to recent progress in lattice calculations
\cite{ FlavourLatticeAveragingGroupFLAG:2024oxs}.
However, the dominant theoretical uncertainty in $\epsK$ arises
from $\Vcb$ \cite{Jwa:2024xjq}.
%

%----------------
% the BGS method
%----------------
Traditionally, the short-distance QCD corrections ($\eta_i$) entering
the master formula for $\epsK$ are evaluated assuming the charm–top
($c–t$) unitarity relation, with subdominant uncertainties arising from
$\eta_{ct}$ and $\eta_{cc}$, both known at next-to-next-to-leading
order (NNLO) in perturbation theory \cite{ Bailey:2015tba,
  Buras2013:EurPhysJC.73.2560, Brod2011:PhysRevLett.108.121801}.
More recently, Brod, Gorbahn, and Stamou (BGS) proposed an alternative
formulation based on the up–top ($u–t$) unitarity relation \cite{
  Brod:2019rzc}, which can potentially reduce these subdominant
uncertainties.
For a pedagogical discussion of these approaches, see Ref.~\cite{
  FlavourLatticeAveragingGroupFLAG:2024oxs}.
%

%--------------------------------
% why is this paper interesting?
%--------------------------------
In this paper, we report updated results for $\epsK$ obtained using
both the traditional $c–t$ unitarity formulation and the BGS method,
incorporating the latest lattice QCD inputs and experimental data.
To highlight the key findings upfront:
\begin{itemize}
\item Using exclusive $\Vcb$ and the traditional method, we observe a
  significant tension—in the range of $4.6\sigma$ to
  $5.2\sigma$—between the SM prediction and the experimental value of
  $\epsK$.\footnote{Here the $5.2\sigma$ tension is obtained with the
  estimate of RBC-UKQCD for $\xi_\text{LD}$, while the $4.6\sigma$
  tension is obtained with the BGI estimate. We use the FNAL/MILC-22
  results for exclusive $\Vcb$. For more details, refer to Sections
  \ref{sec:result} and \ref{sec:conclude}. }
\item Adopting the BGS method increases this tension slightly to
  $4.9\sigma$ to $5.5\sigma$.\footnote{Here the $5.5\sigma$ tension is
obtained with the estimate of RBC-UKQCD for $\xi_\text{LD}$, while the
$4.9\sigma$ tension is obtained with the BGI estimate. We use the
FNAL/MILC-22 results for exclusive $\Vcb$. For more details, refer to
Sections \ref{sec:result} and \ref{sec:conclude}. }
\item In contrast, if we use inclusive $\Vcb$ determined via heavy
  quark expansion, this tension disappears.
\end{itemize}
This points to a deep connection between the $\epsK$ tension and the
long-standing discrepancy between exclusive and inclusive
determinations of $\Vcb$, which itself is at the $3.6\sigma$ to
$4.0\sigma$ level \cite{ Bailey:2018feb, Bailey:2015tba}.\footnote{
For more details, refer to Sections \ref{ssec:Vcb} and
\ref{sec:result}. }

%-----------------------------------------------------------
% main goal : what do we want to achieve out of this paper?
%-----------------------------------------------------------
The main objective of this paper is to present the most up-to-date
Standard Model predictions of $|\epsK|$, incorporating state-of-art
lattice QCD and experimental inputs.
Preliminary results have appeared in Refs.~\cite{ Jwa:2024xjq,
  Jwa:2023uuz, Lee:2023lxz, Lee:2021crz, Kim:2019vic,Bailey:2018aks}.

The remainder of this paper is organized as follows:
\begin{itemize}
\item In Sec.~\ref{sec:rev}, we review the master formulas for $\epsK$
  based on both the traditional $c–t$ and the BGS $u–t$ unitarity
  approaches.
\item Sec.~\ref{sec:input} discusses the determination of input
  parameters from lattice QCD and experiment.
\item In Sec.~\ref{sec:result}, we present our results for $|\epsK|$,
  including a detailed analysis of various input combinations and
  error budgets.
\item We conclude in Sec.~\ref{sec:conclude}.
\end{itemize}

%
% \wlee{\\ EDIT by wlee \\}
%

%% file: master.tex
%
% master.tex
%
\section{Review of $\epsK$}
\label{sec:rev}

We have two kinds of the master formula for $\epsK$: one is the
\texttt{traditional} method of using $\eta_i$ of $c-t$ unitarity
\cite{ Bailey:2015tba, Bailey:2018feb}, and the other is the
\texttt{BGS} method of using $\eta_i$ of $u-t$ unitarity \cite{
  Brod:2019rzc}.
The former method of $c-t$ unitarity is explained in Subsection
\ref{ssec:master-tr-ctu} and the latter method of $u-t$ unitarity
is described in Subsection \ref{ssec:BGS}.

\subsection{Master Formula for $\epsK$ (traditional method)}
\label{ssec:master-tr-ctu}
In the standard model (SM), the direct CP violation parameter $\epsK$
in the neutral kaon system can be re-expressed in terms of the
well-known SM parameters as follows,
\begin{align}
  \epsK
  =& e^{i\theta} \sqrt{2}\sin{\theta} 
  \Big( C_{\eps} X_\text{SD} \hat{B}_{K} 
  + \frac{ \xi_{0} }{ \sqrt{2} } + \xi_\text{LD} \Big) \CL
   &+ \mathcal{O}(\omega \eps^\prime)
   + \mathcal{O}(\xi_0 \Gamma_2/\Gamma_1) \,.
  \label{eq:epsK_SM_0}
\end{align}
This is the master formula, and its derivation is well explained in
Ref.~\cite{ Bailey:2015tba, Bailey:2018feb}.
Here we adopt the same notation and convention as in Ref.~\cite{
  Bailey:2015tba, Bailey:2018feb}.

\subsubsection{Short Distance Contribution to $\epsK$}
\label{sssec:SD}
In the master formula of Eq.~\eqref{eq:epsK_SM_0}, the dominant
leading-order effect ($\approx +107\%$) comes from the short distance
(SD) contribution proportional to $\BK$.
Here, $C_{\eps}$ is a dimensionless parameter defined as:
\begin{align}
  C_{\eps}
  &\equiv \frac{ G_{F}^{2} F_K^{2} m_{K^{0}} M_{W}^{2} }
  { 6\sqrt{2} \pi^{2} \Delta M_{K} } \cong 3.63 \times 10^{4} \,,
  \label{eq:C_eps}
\end{align}
Here, $X_\text{SD}$ represents the short distance effect from the
Inami-Lim functions of the box diagrams \cite{ Inami:1980fz}:
\begin{align}
  X_\text{SD} &\equiv
  \Im\lambda_t \Big[ \Re\lambda_c \eta_{cc} S_0(x_c)
    -\Re\lambda_t \eta_{tt} S_0(x_t) \CL & \quad - (\Re\lambda_c -
    \Re\lambda_t) \eta_{ct} S_0(x_c,x_t) \Big] 
  \label{eq:X_SD}
  \\
  &\cong 6.24 \times 10^{-8} \,,
\end{align}
where $\lambda_i = V^*_{is} V_{id}$ is a product of the CKM matrix
elements with $i = u,c,t$, $x_i = m_i^2/M_W^2$.
Here $\eta_{i}$ with $i = cc, ct, tt$ represent the QCD corrections of
higher order in $\alpha_s$ \cite{ Herrlich1996:NuclPhysB.476.27}.
We call this formula the \texttt{traditional} method for $\eta_i$ of
$c-t$ unitarity, as in Ref.~\cite{ Brod:2019rzc}.
Here, $m_i = m_i(m_i)$ is the scale invariant $\MSb$ quark mass.
There exists a potential issue with poor convergence of perturbation
theory for $\eta_{cc}$ at the charm scale, which is discussed properly
in Ref.~\cite{ Bailey:2015tba}.
Here, $S_0$'s are Inami-Lim functions which are defined in Ref.~\cite{
  Inami:1980fz}, and Refs.~\cite{ Bailey:2018feb, Bailey:2015tba}.
In $X_\text{SD}$ of Eq.~\eqref{eq:X_SD}, the $S_0(x_t)$ term from the
top-top contribution in the box diagrams describes about
$+71\%$ of $X_\text{SD}$, the $S_0(x_c, x_t)$ term from the
top-charm contribution takes over about $+48\%$ of
$X_\text{SD}$, and the $S_0(x_c)$ term from the charm-charm
contribution depicts about $-19\%$ of $X_\text{SD}$.
%
%+71.08, +47.64, -18.72

Here, the kaon bag parameter $\BK$ is defined as
\begin{align}
  \BK &\equiv B_K(\mu) b(\mu) \cong 0.76 \,,
  \\
  B_K(\mu)
  &\equiv
  \frac{ \mate{ \bar{K}^{0} }{ O_{LL}^{\Delta S = 2}(\mu) }{ K^{0} } }
       { \frac{8}{3}
         \mate{\bar{K}^0}{\bar{s} \gamma_{\mu}\gamma_5 d}{0}
         \mate{0}{\bar{s} \gamma^{\mu}\gamma_5 d}{K^0} } \CL
  &= \frac{ \mate{\bar{K}^0}{O_{LL}^{\Delta S = 2}(\mu)}{K^0} }
       { \frac{8}{3} F_K^2 m_{K^0}^2 } \,,
  \\
  O_{LL}^{\Delta S = 2}(\mu)
  &\equiv [\bar{s}\gamma_{\mu}(1-\gamma_5)d]
         [\bar{s}\gamma^{\mu}(1-\gamma_5)d] \,,  
\end{align}
where $b(\mu)$ is the renormalization group (RG) running factor to
make $\BK$ invariant with respect to the renormalization scale and
scheme:
\begin{align}
  b(\mu) &= [ \alpha_{s}^{(3)}(\mu) ]^{-2/9} K_+(\mu) \,.
\end{align}
Here, details on $K_+(\mu)$ are given in Ref.~\cite{Bailey:2015tba}.

\subsubsection{Long Distance Contribution to $\epsK$}
\label{sssec:LD}
There are two kinds of long distance (LD) contributions on $\epsK$:
one is the absorptive LD effect from $\xi_0$ and the other is the
dispersive LD effect from $\xi_\text{LD}$.
The absorptive LD effects are defined as
\begin{align}
  \tan{\xi_0} &\equiv \frac{\Im A_0}{\Re A_0} \,,
  \\
  \tan{\xi_2} &\equiv \frac{\Im A_2}{\Re A_2} \,.
\end{align}
They are related with each other through $\eps^\prime$:
\begin{align}
  \eps^\prime
  & \equiv 
  e^{i(\delta_2-\delta_0)} \frac{i\omega}{\sqrt{2}} 
  \Big( \tan{\xi_2} - \tan{\xi_0} \Big) 
  \nonumber \\
  & =   e^{i(\delta_2-\delta_0)} \frac{i\omega}{\sqrt{2}} 
  ( \xi_2 - \xi_0 ) + \mathcal{O}(\xi^3_i) \,.
  \label{eq:epsp}
\end{align}
The overall contribution of the $\xi_0$ term to $\epsK$ is about
$-9.4\%$.
%jwa
%indirect xi_0: -1.738/22.28=-0.07801
%direct xi_0: -2.102/22.28=-0.09434

The dispersive LD effect is defined as
\begin{align}
	\xi_\text{LD} &=  \frac{m^\prime_\text{LD}}{\sqrt{2} \Delta M_K} \,,
\label{eq:xiLD}
\end{align}
where
\begin{align}
  \label{eq:mLD}
  m^\prime_\text{LD}
  &= -\Im \left[ \mathcal{P}\sum_{C} 
    \frac{\mate{\wbar{K}^0}{H_\text{w}}{C} \mate{C}{H_\text{w}}{K^0}}
         {m_{K^0}-E_{C}}  \right]
  \,.
\end{align}
if the CPT invariance is well respected.
The overall contribution of the $\xi_\text{LD}$ to $\epsK$ is about
$\pm 2.2\%$.
% 1.738/22.28*0.4/math.sqrt(2)=0.02206

\subsection{Master Formula for $\epsK$ (BGS method)}
\label{ssec:BGS}
The BGS method was reported in Ref.~\cite{ Brod:2019rzc}.
The original idea on $u-t$ unitarity was introduced in Ref.~\cite{
  Christ:2012se} in order to find a better method to compute weak
matrix elements on the lattice with dynamical charm quarks.
There has been a number of speculations on the better convergence in
the perturbative expansion with respect to the charm quark
contribution, as in Ref.~\cite{ 10.1093/oso/9780198855743.001.0001}.
A pedagogical review on the $c-t$ unitarity and the $u-t$ unitarity is
available in FLAG-24 \cite{ FlavourLatticeAveragingGroupFLAG:2024oxs}.

The BGS master formula for $\epsK$ is
\begin{align}
  \epsK
  =& e^{i\theta} \sqrt{2}\sin{\theta} 
  \Big( C_{\eps} X^\text{BGS}_\text{SD} \hat{B}_{K} 
  + \frac{ \xi_{0} }{ \sqrt{2} } + \xi_\text{LD} \Big) \CL
   &+ \mathcal{O}(\omega \eps^\prime)
   + \mathcal{O}(\xi_0 \Gamma_2/\Gamma_1) \,.
  \label{eq:epsK_BGS_0}
\end{align}
Here note that the short-distance contribution term $X_\text{SD}$ in
Eq.~\eqref{eq:epsK_SM_0} and Eq.~\eqref{eq:X_SD} in the traditional
method of $c-t$ unitarity is replaced by the new term
$X^\text{BGS}_\text{SD}$ in the BGS method of $u-t$ unitarity.
\begin{align}
  X^\text{BGS}_\text{SD} &= \Vcb^2 \lambda^2 \bar{\eta} \;
  \Big[ \Vcb^2 (1-\bar{\rho}) \, \eta_{tt}^\text{BGS} \, \CS_{tt}( x_c, x_t)
    \nonumber \\
   & \hspace{7em}  - \eta_{ut}^\text{BGS} \, \CS_{ut}( x_c, x_t) \Big]
  \label{eq:X-SD-BGS-1}
\end{align}
Here the QCD corrections $\eta_i$ of $u-t$ unitarity \cite{
  Brod:2019rzc} are
\begin{align}
  \eta_{tt}^\text{BGS} &= \eta_{tt}^\text{NLL}
  = 0.55(1 \pm 4.2\% \pm 0.1\%)
  \label{eq:eta:BGS:tt-1}
  \\
  \eta_{ut}^\text{BGS} &= \eta_{ut}^\text{NNLL}
  = 0.402 (1 \pm 1.3\% \pm 0.2\% \pm 0.2\%)
  \label{eq:eta:BGS:ut-1}
\end{align}
The modified Inami-Lim functions are
\begin{align}
  \CS_{tt}(x_c, x_t) &= S(x_t) + S(x_c) - 2 S(x_c, x_t)
  \\
  \CS_{ut}(x_c, x_t) &= S(x_c) - S(x_c, x_t)
\end{align}
Here $C_{\eps}$, $\xi_{0}$, $\xi_\text{LD}$ are the same as in the
previous Subsection \ref{ssec:master-tr-ctu}.
%

%% file: input.tex
%
% input.tex
%
\section{Input Parameters}
\label{sec:input}
We need values of 19 input parameters defined in the standard model
(SM) in order to evaluate $\epsK$ directly from the SM.
Out of the 19 parameters, it is possible to obtain 9 parameters such
as $\BK$, $\Vcb$, $\xi_0$, $\xi_2$, $\xi_\text{LD}$, $\Vus$, $\Vud$,
$F_K$, and $m_c(m_c)$ directly from lattice QCD.
Here, we describe how to obtain the entire input parameters from
lattice QCD results and from experimental results in detail.
%

%------------------
% input parameters
%------------------

\input{ ip_WP.tex}

\input{ ip_Vcb.tex}

\input{ ip_BK.tex}

\input{ ip_xi0.tex}

\input{ ip_xi_LD.tex}

\input{ ip_m_t.tex}

\input{ ip_m_c.tex}

\input{ ip_W_mass.tex}

\input{ ip_eta.tex}

\input{ ip_other.tex}

%% file: ip_WP.tex
%
% ip_WP.tex
%
\subsection{Wolfenstein Parameters }
\label{ssec:wp}
The CKMfitter \cite{ ValeSilva:2024jml} and UTfit \cite{ Bona:2024bue,
  Bona:2006ah} collaborations keep updating the Wolfenstein parameters
(WP) \cite{ Winstein1993:RevModPhys.65.1113} ($\lambda$, $A$,
$\bar{\rho}$, $\bar{\eta}$) on their web sites, which are determined
by the global unitarity triangle (UT) fit.
Results of the CKMfitter (2024) and the UTfit (2024) are summarized in
Table \ref{tab:wp}.

\input{tab_wp}

As pointed out in Ref.~\cite{ Bailey:2015tba}, the Wolfenstein
parameters extracted by the global UT fit have unwanted correlation
with $\epsK$, because $\epsK$ is used as an input to obtain them.
Hence, in order to avoid this correlation, we take another set of the
Wolfenstein parameters determined from the angle-only-fit (AOF)
suggested in Ref.~\cite{ Bevan2013:npps241.89}.
In the AOF, $\epsK$, $\BK$, and $\Vcb$ are not used as inputs to
determine the UT apex ( $\bar{\rho}$, $\bar{\eta}$).
We present the AOF results for WP in Table \ref{tab:wp}.

We can determine $\lambda$ from $\Vus$ and $\Vud$, or from $r$, using
Eq.~\eqref{eq:lambda-1}.
\begin{align}
  \lambda &= \frac{ \Vus }{ \sqrt{ \Vud^2 + \Vus^2 } }
  = \frac{ r }{ \sqrt{ 1 + r^2 } } \,,
  \label{eq:lambda-1}
  \\
  r &= \frac{\Vus}{\Vud} \,.
  \label{eq:lambda-2}
\end{align}
Both results for $\Vus$ and $\Vud$, and $r$ are obtained from lattice
QCD \cite{ FlavourLatticeAveragingGroupFLAG:2024oxs}.
In Table \ref{tab:input-Vus-Vud} \subref{tab:Vus+Vud}, we summarize
results for $\Vus$ and $\Vud$.
In Table \ref{tab:input-Vus-Vud} \subref{tab:Vus/Vud}, we present
results for $r$.
Here, we determine $\lambda$ from $r$ in Table \ref{tab:input-Vus-Vud}
\subref{tab:Vus/Vud}, since its error is less than the other
combination ($\Vus$ and $\Vud$).
We present results for $\lambda$ determined from $r$ in the AOF column
of Table \ref{tab:wp}.
The Wolfenstein parameter $A$ is determined directly from $\Vcb$,
which will be discussed later in Section \ref{ssec:Vcb}.
\input{tab_Vud_Vus}
Recently the UTfit collaboration has updated values for the WP in
Ref.~\cite{ Bona:2024bue}.
Here, we use the results of AOF in Table \ref{tab:wp} to evaluate
$|\epsK|$, in order to avoid unwanted correlation between ( $\eps_K$,
$\BK$, $\Vcb$) and ( $\bar\rho$, $\bar\eta$).

%
% \wlee{\\ EDIT by wlee \\}
%

%% file: tab_wp.tex
\begin{table*}
  \caption{Wolfenstein parameters (WP). Both CKMfitter and UTfit
    groups use the global unitarity triangle fit. Here, AOF represents
    the angle-only-fit.}
  \label{tab:wp}
  \renewcommand{\arraystretch}{1.3}
  \begin{ruledtabular}
    \begin{tabular}{@{\quad} c @{\quad}| l c @{\quad}| l c @{\quad}| l c }
      WP
      & \multicolumn{2}{c|}{CKMfitter}
      & \multicolumn{2}{c|}{UTfit}
      & \multicolumn{2}{c}{AOF} \\ \hline
      $\lambda$
      & $0.22498^{+0.00023}_{-0.00021}$ & \cite{ ValeSilva:2024jml}
      & \blue{ 0.22519(83) }            & \cite{ UTfit:2022hsi, Bona:2006ah}
      %& \red{  0.22597(43) }            & \cite{ FLAG:2024}
      & \red{  0.22536(42) }            & \cite{ FlavourLatticeAveragingGroupFLAG:2024oxs}
      \\ \hline
      $\bar{\rho}$
      & $0.1562^{+0.0112}_{-0.0040}$ & \cite{ ValeSilva:2024jml}
      & \blue{ 0.160(9) }            & \cite{ Bona:2024bue, Bona:2006ah}
      & \red{ 0.159(16) }            & \cite{ Bona:2024bue}
      \\ \hline
      $\bar{\eta}$
      & $0.3551^{+0.0051}_{-0.0057}$ & \cite{ ValeSilva:2024jml}
      & \blue{ 0.346(9) }            & \cite{ Bona:2024bue, Bona:2006ah}
      & \red{ 0.339(10) }            & \cite{ Bona:2024bue}
    \end{tabular}
  \end{ruledtabular}
\end{table*}

%% file: tab_Vud_Vus.tex
\begingroup
\squeezetable
\begin{table}[h!]
  \caption{ CKM matrix elements: \subref{tab:Vus+Vud} $\Vus$ and
    $\Vud$, and \subref{tab:Vus/Vud} $r
    = \Vus/\Vud$. } \label{tab:input-Vus-Vud}
  \subtable[$\Vus$ and $\Vud$]{
    \renewcommand{\arraystretch}{1.2}
    \hspace*{-4mm}
    \begin{ruledtabular}
      \small
      \begin{tabular}{ l l l l }
        type  &  $\Vus$  & $\Vud$  & Ref.
        \\ \hline
        Lattice  & \multirow{2}{*}{ 0.22483(61) }
                 & \multirow{2}{*}{ 0.97439(14) }
                 & \multirow{2}{*}{ FLAG-24
                 \cite{ FlavourLatticeAveragingGroupFLAG:2024oxs} }%p79t20
        \\
        $N_f = 2+1+1$ & & &
        \\ \hline
        Lattice  & \multirow{2}{*}{ \blue{0.22481(58)} }
                 & \multirow{2}{*}{ \blue{0.97440(13)} }
                 & \multirow{2}{*}{ FLAG-24
                 \cite{ FlavourLatticeAveragingGroupFLAG:2024oxs} }%p79t20
        \\
        $N_f = 2+1$ & & &
        \\ \hline
        nuclear  & \multirow{2}{*}{ 0.2277(13) }
                 & \multirow{2}{*}{ 0.97373(31) }
                 & \multirow{2}{*}{ PDG-22 \cite{ Workman:2022ynf} }
        \\
        $\beta$ decay & & &
      \end{tabular}
    \end{ruledtabular}  
    \label{tab:Vus+Vud}
  } % subtable
  %
  %\vfill
  %
  \subtable[$r = \Vus/\Vud$]{
    \renewcommand{\arraystretch}{1.65}
    \hspace*{-4mm}
    \begin{ruledtabular}
      \small
      \begin{tabular}{ l l c }
        type  &  $r$  & Ref.
        \\ \hline
          %
        %QCD     & \blue{ 0.22328(58) } & FLAG-24 \cite{FLAG:2024}
      $f_{K^{\pm}}/f_{\pi^{\pm}}$     & \blue{ 0.23126(50) } & FLAG-24 \cite{FlavourLatticeAveragingGroupFLAG:2024oxs}%p75
        \\
        %QCD+QED & \red{  0.23197(47) } & FLAG-24 \cite{FLAG:2024}
      $f_K/f_\pi$ & \red{  0.23131(45) } & FLAG-24 \cite{FlavourLatticeAveragingGroupFLAG:2024oxs}%p76
          \\
      \end{tabular}
    \end{ruledtabular}
    \label{tab:Vus/Vud}
  } %%% \subtable
\end{table}
\endgroup

%% file: ip_Vcb.tex
%
% ip_Vcb.tex
%
\subsection{$\Vcb$ }
\label{ssec:Vcb}
In Table \ref{tab:Vcb} we summarize updated results for both
exclusive $\Vcb$ and inclusive $\Vcb$.
Here the \texttt{ex-comb} means that the results for $\Vcb$ are
obtained using the combined fitting of multiple decay channels such as
$\Vcb$ determined from the $\BtoDst$ decays, $\Vcb$ determined from
the $\BtoD$ decays, and constraints by $\Vub/\Vcb$ determined from the
ratio of $\RBF(\Lambda_b)$ or $\RBF(B_s)$.
\begin{align}
  \RBF(\Lambda_b) &= \frac{ \cB ( \Lbtop ) }{ \cB ( \LbtoLc ) }
  \\
  \RBF(B_s) &= \frac{ \cB ( \BstoK ) }{ \cB ( \BstoDs ) }
\end{align}
%

%----------------
% FNAL/MILC 2022
%----------------
In Ref.~\cite{ FermilabLattice:2021cdg} the FNAL/MILC collaboration
reported results for exclusive $\Vcb$ determined using the BGL
parametrization methods applied to both the Belle and BaBar
experimental results combined with the semileptonic form factors
obtained using the lattice QCD tools for the $\BtoDst$ decays at both
zero ($w = 0$) and non-zero ($w \ne 0$) recoil points.
The semileptonic form factors are extracted from the 2pt and 3pt
correlation function data measured on the MILC asqtad ensembles \cite{
  Bazavov:2009bb, Aubin:2004wf, Bernard:2001av}.
The FNAL/MILC results for $\Vcb$ are summarized in Table \ref{tab:Vcb}
\subref{tab:ex-Vcb}.

%-----------
% FLAG 2024
%-----------
The FLAG adopted the \texttt{ex-comb} method to obtain their final
results in Ref.~\cite{ FlavourLatticeAveragingGroupFLAG:2024oxs}.
In their version of \texttt{ex-comb} they use results from three
channels: (1) results for $\Vcb$ determined using the BGL
parametrization method for the $\BtoDst$ decay modes from both Belle
and BaBar, and (2) results for $\Vcb$ determined using the BCL
parametrization method for the $\BtoD$ decay modes from both Belle and
BaBar combined with the semileptonic form factors obtained using the
lattice QCD tools by the FNAL/MILC \cite{MILC:2015uhg} and
HPQCD collaborations \cite{Na:2015kha}, and (3) results for
$\Vub/\Vcb$ determined from the ratio of $\RBF(B_s)$
\cite{McLean:2019qcx} excluding results for $\RBF(\Lambda_b)$
\cite{HFLAV:2019otj}.
Here the caveats are that they (FNAL/MILC and HPQCD) share the same
asqtad gauge configuration ensembles of MILC \cite{MILC:2009mpl} to perform
the measurements, which leads to unwanted correlation between their
results.
They (FLAG) just neglected it.
The FLAG results for $\Vcb$ are summarized in Table \ref{tab:Vcb}
\subref{tab:ex-Vcb}.
%

%------------
% HFLAV 2023
%------------
Recently the HFLAV collaboration reported the \texttt{ex-comb} results
for $\Vcb$ in Ref.~\cite{HFLAV:2022esi}.
In their version of \texttt{ex-comb}, they use results of multiple
channels: (1,2) results for $\Vcb$ obtained using the CLN
parametrization methods applied to the experimental results for the
$\BtoDst$ and $\BtoD$ decay modes combined with the semileptonic form
factors determined using the lattice QCD tools by FLAG 2021
\cite{FlavourLatticeAveragingGroupFLAG:2021npn} and FNAL/MILC
\cite{FermilabLattice:2014ysv}, and (3) results for exclusive $\Vcb$
obtained using the CLN and BGL methods to analyze the experimental
results of LHCb for the $\BstoDs$ and $\BstoDsst$ decays combined with
the semileptonic form factors determined from lattice QCD by HPQCD
\cite{McLean:2019sds, McLean:2019qcx}, and (4) results for exclusive
$\Vub/\Vcb$ determined from the ratio of $\RBF(\Lambda_b)$ and
$\RBF(B_s)$ with the form factors obtained from lattice QCD by
Ref.~\cite{Detmold:2015aaa} and HPQCD \cite{McLean:2019sds}.
Here the caveats are that they also neglect unwanted correlation among
the semileptonic form factors determined from lattice QCD due to
sharing the MILC gauge ensembles in the data measurements.
The HFLAV results for $\Vcb$ are collected in Table \ref{tab:Vcb}
\subref{tab:ex-Vcb}.
%

%------------
% HPQCD 2023
%------------
In Ref.~\cite{Harrison:2023dzh} the HPQCD collaboration reported updated
\texttt{ex-comb} results for exclusive $\Vcb$.
In their version of \texttt{ex-comb} they use results from two
independent channels: (1) results for $\Vcb$ determined from the Belle
untagged data of the $\BtoDst$ decays ~\cite{Belle:2018ezy} and (2) results
for $\Vcb$ obtained from the LHCb untagged data of $\BstoDsst$ decays
\cite{LHCb:2020hpv} combined with the semileptonic form factors obtained
using lattice QCD tools by the HPQCD collaboration.
They adopted the heavy-HISQ method \cite{Harrison:2021tol} to obtain the
semileptonic form factors, in which they use the HISQ action (one of
the improved staggered quark actions) to interpolate to the charm
quark mass and to extrapolate to the bottom quark mass.
They use the MILC HISQ gauge ensembles \cite{Follana:2006rc} for the numerical
study.
Their results are consistent with those of FNAL/MILC 2022, but just
have bigger errors.
The HPQCD results for $\Vcb$ are collected in Table \ref{tab:Vcb}
\subref{tab:ex-Vcb}.
%
%---------------
% table of V_cb
%---------------
\input{tab_Vcb}

%----------------
% inclusive V_cb
%----------------
In Table \ref{tab:Vcb} \subref{tab:ex-Vcb} we summarize results for
inclusive $\Vcb$.
The tension between inclusive $\Vcb$ and exclusive $\Vcb$ remains in
the $3.1\sigma \sim 4.0\sigma$ level as before.
\input{fig_Vcb_CLN_BGL.tex}
One good news is that the difference in exclusive $\Vcb$ between the
CLN\footnote{ For more details, refer to Refs.~\cite{ Caprini:1995wq,
  Caprini:1997mu}. }and BGL\footnote{ For more details, refer to
Ref.~\cite{Boyd:1995sq}. } parametrization methods has disappeared
since 2019 \cite{Kim:2019vic}, as one can see in
Fig.~\ref{fig:Vcb-cln-bgl-1}.
In Fig.~\ref{fig:Vcb-cln-bgl-1}, the substantial discrepancy between
the exclusive $\Vcb$ values obtained using the CLN paramtrization 
and those obtained using the BGL parametrization were observed until the
first version of the FLAG 2019 report
\cite{FlavourLatticeAveragingGroup:2019iem} (refer to the magenta
label \magenta{FLAG-19} in Fig.~\ref{fig:Vcb-cln-bgl-1}).
This discrepancy has disappeared since the SWME 2019 report
\cite{Kim:2019vic} (refer to the green label \greenfour{SWME-19}).
The critical issues on the debates have been resolved by the time
evolution as in Fig.~\ref{fig:Vcb-cln-bgl-1}.
Details on the previous debates and unresolved issues between CLN and BGL
are described in \cite{Bailey:2018feb}.
\input{tab_Vcb_CLN_BGL.tex}
%

%
%\wlee{\\ EDIT by wlee \\}
%

%% file: tab_Vcb.tex
\begin{table}[h]
%%  \footnotesize
  \renewcommand{\arraystretch}{1.4}
  \caption{Results for $\Vcb$ in units of $1.0\times 10^{-3}$.}
  \label{tab:Vcb}
  \subtable[Exclusive $\Vcb$]{
    \hspace*{-4mm}
    \begin{ruledtabular}
      \begin{tabular}{l l l l}
        channel & value & Ref. & source \\ \hline
%---------------------------
% FNAL/MILC (BELLE + BABAR) 
%---------------------------        
        $B\to D^* \ell \bar{\nu}$
        & \red{38.40(78)}
        % & BGL
        & \cite{ FermilabLattice:2021cdg} % p27e76
        & FNAL/MILC-22
        \\
%----------
% FLAG 
%----------
        ex-comb
        %& \red{39.53(54)}
        & \red{39.46(53)}
        % & comb
        %& \cite{ FLAG:2024} % p177e283
        & \cite{ FlavourLatticeAveragingGroupFLAG:2024oxs} % p181e282
        & FLAG-24
        \\
%----------
% HFLAV
%----------
        ex-comb & \red{39.10(50)}
        % & comb
        & \cite{ HFLAV:2022esi} % p120e221
        & HFLAV-23
        \\
%----------
% HPQCD
%----------
        ex-comb & \red{39.03(56)(67)}
        % & comb
        & \cite{ Harrison:2023dzh} % p22e51
        & HPQCD-23
      \end{tabular}
    \end{ruledtabular}
    \label{tab:ex-Vcb} 
  } %%% \subtable
  \subtable[Inclusive $\Vcb$]{
    \hspace*{-4mm}
    \begin{ruledtabular}
      \small
      \begin{tabular}{l l c l}
        channel & value & Ref. & source \\ \hline
        kinetic scheme & \blue{ 42.16(51) }
        & \cite{ Bordone:2021oof}
        & Gambino-21
        \\ 
        1S scheme      & \red{ 41.98(45) }
        & \cite{ HFLAV:2022esi}
        & 1S-23
      \end{tabular}
    \end{ruledtabular}
    \label{tab:in-Vcb} 
  } %%% \subtable
\end{table}

%% file: fig_Vcb_CLN_BGL.tex
\begin{figure}[htb]
  \includegraphics[width=1.0\linewidth]{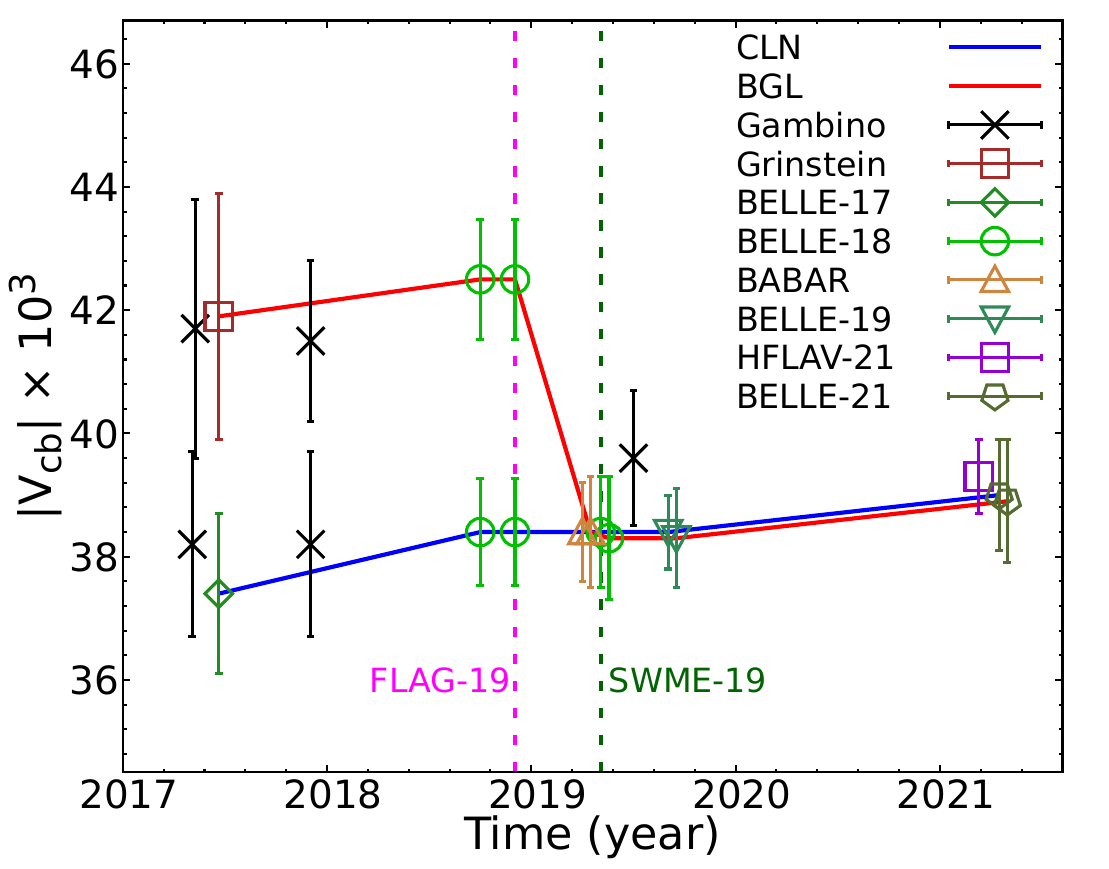}
  \caption{ Comparison of $\Vcb$ [in units of $1.0 \times 10^{-3}$]
    obtained using the CLN (blue line) and BGL (red line)
    parametrization methods. Details on the legends such as BELLE-17
    are presented in Table \ref{tab:Vcb-cln-bgl-1}. }
  \label{fig:Vcb-cln-bgl-1}
\end{figure}

%% file: tab_Vcb_CLN_BGL.tex
\begin{table}[htbp]
  \renewcommand{\arraystretch}{1.4}
  \caption{Results of exclusive $\Vcb$ [in units of $1.0\times
      10^{-3}$] obtained using the CLN and BGL parametrization
    methods. Here the time is given in the format of
    \texttt{year-month}.}
  \label{tab:Vcb-cln-bgl-1}
  \begin{ruledtabular}
    \small
    \begin{tabular}{l l l l l}
      label ID &  time & method & value & Ref.  \\ \hline
%%%%
% PLB769 Gambino
%%%%
      Gambino & \multirow{2}{*}{2017-04} & CLN  & 38.2(15)  & \cite{Bigi:2017njr} 	\\
      Gambino &	 		         & BGL  & 41.7(21) & \cite{Bigi:2017njr}  	\\ \hline
%%%%
% 2017 BELLE
%%%%
      BELLE-17  & \multirow{2}{*}{2017-05} & CLN  & 37.4(13)  &\cite{Belle:2017rcc} \cite{Grinstein:2017nlq} \\
      Grinstein &  			   & BGL  & 41.9(20) & \cite{Grinstein:2017nlq}  	\\ \hline
%%%%
% JHEP1711 Gambino
%%%%
      Gambino & \multirow{2}{*}{2017-11} & CLN  & 38.2(15)  & \cite{Bigi:2017jbd} 	\\
      Gambino &	  		   & BGL  & 41.5(13) & \cite{Bigi:2017jbd}  	\\ \hline
%%%%
% 2018 BELLE v1
%%%%
      BELLE-18(v1) & \multirow{2}{*}{2018-09} & CLN  & 38.4(9)  & \cite{Belle:2018ezy}  \\
      BELLE-18(v1) &				& BGL  & 42.5(10) & \cite{Belle:2018ezy}  \\ \hline
%%%%
% 2018 BELLE v2
%%%%
      BELLE-18(v2) & \multirow{2}{*}{2018-11} & CLN  & 38.4(9)  & \cite{Belle:2018ezy}  \\
      BELLE-18(v2) & 			      & BGL  & 42.5(10) & \cite{Belle:2018ezy}  \\ \hline
%%%%
% 2019 BABAR 
%%%%
      BABAR-19 & \multirow{2}{*}{2019-03} & CLN  & 38.4(8)  & \cite{BaBar:2019vpl}  \\
      BABAR-19 &			    & BGL  & 38.4(9)  & \cite{BaBar:2019vpl}  \\ \hline
%%%%
% 2018 BELLE v3
%%%%
      BELLE-18(v3) & \multirow{2}{*}{2019-04} & CLN  & 38.4(9)  & \cite{Belle:2018ezy}  \\
      BELLE-18(v3) &		              & BGL  & 38.3(10) & \cite{Belle:2018ezy}  \\ \hline
%%%%
% 2019 Gambino
%%%%
      Gambino	&		2019-05 & BGL  & 39.6(11) & \cite{Gambino:2019sif} \\ \hline
%%%%
% 2021 HFLAV
%%%%
      HFLAV-21 	&		2021-04 & CLN  & 39.3(6)  & \cite{HFLAV:2019otj}   \\ \hline
%%%%
% 2021 BELLE
%%%%
      BELLE-21 & \multirow{2}{*}{2021-04} & CLN  & 39.0(9)  & \cite{Belle:2018ezy}   \\
      BELLE-21 &  			& BGL  & 38.9(10) & \cite{Belle:2018ezy}
    \end{tabular}
  \end{ruledtabular}
\end{table}

%% file: ip_BK.tex
%
% ip_BK.tex
%
\subsection{$\BK$}
\label{ssec:bk}
In the FLAG review 2024 \cite{
  FlavourLatticeAveragingGroupFLAG:2024oxs}, they report lattice QCD
results for $\BK$ with $N_f=2$, $N_f=2+1$, and $N_f= 2+1+1$.
Here we prefer the $\BK$ results with $N_f=2+1$ to those with $N_f=
2+1+1$.
The physical reasons for this choice are given in our previous paper
Ref.~\cite{ Bailey:2018feb}.
Basically the master formula for $\epsK$ is derived for $N_f=2+1$
but not for $N_f= 2+1+1$.
Correspondingly the $\BK$ parameter is also defined for $N_f=2+1$.
Hence, it is significantly more convenient and straightforward to use
the $\BK$ results with $N_f=2+1$.

Even though a theoretical procedure to incorporate the charm quark
effect (with $N_f = 2+1+1$) systematically on the $\epsK$ calculation
is available in Refs.~\cite{ Christ:2012se, Bai:2014cva}, this method
is in the premature stage of exploratory study, and far away from the
precision measurement yet.
There has been an interesting attempt to combine all the lattice
results for $\BK$ with $N_f=2+1$ and $N_f=2+1+1$ to obtain a result
for $\BK$ with $N_f=2+1$ in Ref.~\cite{ Gorbahn:2024qpe}.

We prefer using the results for $\BK$ with $N_f=2+1$ to those with
$N_f=2$ since the approximation of quenching the vacuum polarization
contributions of the strange quark is an unwanted damage to the
precision measurement.

\input{tab_bk}
In Table \ref{tab:bk} we present the FLAG 2024 results for $\BK$ with
$N_f = 2+1$.
Here they take a global average over the five data points from BMW
11 \cite{ Durr:2011ap}, Laiho 11 \cite{ Laiho:2011np}, RBC/UKQCD 14
\cite{ Blum:2014tka}, SWME 15 \cite{ Jang:2015sla}, and RBC/UKQCD 24
\cite{ Boyle:2024gge}.
The FLAG-24 in the table represents the final combined result for
$\BK$.
Here we use this (FLAG-24) for our evaluation of $\epsK$.

%
%\wlee{\\ EDIT by wlee \\}
%

%% file: tab_bk.tex
\begin{table}[hbtp]
  \caption{$\BK$ in lattice QCD with $N_f = 2+1$.}
  \label{tab:bk}
  \renewcommand{\arraystretch}{1.4}
  \begin{ruledtabular}
    \begin{tabular}{@{\quad} l @{\quad} c @{\quad} l @{\quad}}
      Collaboration & Ref. & $\BK$  \\ \hline
      RBC/UKQCD 24  & \cite{Boyle:2024gge} & $0.7436(25)(78)$   \\
      SWME 15       & \cite{Jang:2015sla}  & $0.735(5)(36)$     \\
      RBC/UKQCD 14  & \cite{Blum:2014tka}  & $0.7499(24)(150)$  \\
      Laiho 11      & \cite{Laiho:2011np}  & $0.7628(38)(205)$  \\
      BMW 11        & \cite{Durr:2011ap}   & $0.7727(81)(84)$  \\ \hline
      %                                  
%%%      FLAG-23       & \cite{FlavourLatticeAveragingGroupFLAG:2021npn}
%%%      & $0.7625(97)$    \\
      FLAG-24       & \cite{FlavourLatticeAveragingGroupFLAG:2024oxs}     & \red{ $0.7533(91)$ } %p96e111
    \end{tabular}
  \end{ruledtabular}
\end{table}

%% file: ip_xi0.tex
%
% ip_xi0.tex
%
\subsection{$\xi_0$ }
\label{ssec:xi0}
The absorptive part of long distance effects in $\epsK$ is
parametrized by $\xi_0$.
We can express $\eps'/\eps$ in terms of $\xi_0$ and $\xi_2$ as
follows,
\begin{align}
  & \xi_0 \equiv \arctan \bigg( \frac{\Im A_0}{\Re A_0} \bigg)
  = \frac{\Im A_0}{\Re A_0} + \mathcal{O}(\xi_0^3)
  \\ 
  & \xi_2 \equiv \arctan \bigg(\frac{\Im A_2}{\Re A_2} \bigg)
  = \frac{\Im A_2}{\Re A_2} + \mathcal{O}(\xi_2^3)
  \\
  & \Re \bigg(\frac{\eps'}{\eps} \bigg) = 
  \frac{\omega}{\sqrt{2} |\epsK|} (\xi_2 - \xi_0) \,.
  \label{eq:e'/e:xi0}
\end{align}
Here $\xi_I$ with isospin $I=0,2$ represents a phase of the amplitude
$A_I$\footnote{The definition of $A_I$ is given in Eq.~(29) of
Ref.~\cite{Bailey:2015tba}}.

There are two independent methods to determine $\xi_0$ in lattice QCD:
one is the indirect method and the other is the direct method.
The indirect method is to determine $\xi_0$ using
Eq.~\eqref{eq:e'/e:xi0} with lattice QCD results for $\xi_2$ combined
with experimental results for $\eps'/\eps$, $\epsK$, and $\omega$.
The direct method is to determine $\xi_0$ directly using lattice QCD
results for $\Im A_0$ combined with experimental results for $\Re
A_0$.
%

%
\input{tab_xi0}
In Table \ref{tab:xi0} \subref{subtab:exp-ReA0-ReA2}, we summarize
experimental results for $\Re A_0$, $\Re A_2$, $\omega$, $|\epsK|$,
and $\Re(\eps'/\eps)$.
In Table \ref{tab:xi0} \subref{subtab:latt-ImA0}, we present
lattice QCD results for $\Im A_0$.
In Table \ref{tab:xi0} \subref{subtab:latt-ImA2}, we present
lattice QCD results for $\Im A_2$.
In Table \ref{tab:xi0} \subref{subtab:xi0}, we report results for
$\xi_0$ obtained, using the indirect and direct methods.
Here we use the results of both the indirect and direct method for
$\xi_0$ to evaluate $\epsK$.
A merit for the indirect method is that its systematic and statistical
errors for $\xi_0$ are much less than those for the direct method.
A merit for the direct method is that the theoretical interpretation
is easy and straightforward, while a disadvantage is that its error is
about 3 times bigger than that of the indirect method.

In Ref.~\cite{ RBC:2020kdj}, the RBC-UKQCD collaboration calculated
amplitudes of $\Im A_0$ and $\Im A_2$ of $K \to \pi\pi$ decays using
the domain wall fermions for light quarks by imposing G-parity
boundary condition (BC) in order to tune the energy of the $I=0,2$ two
pion state very close to the kaon mass.
Recently, the RBC-UKQCD collaboration has calculated the same
amplitudes of $\Im A_0$ and $\Im A_2$ using an alternative approach of
variational methods by imposing periodic BC on domain wall fermions
\cite{ RBC:2023ynh}.
The main advantage of this new approach is that it produces the
correct physical kinematics for the excited states of the two pions
state.
It turns out that the overall error for $\Im A_0$ with the periodic BC
is 1.7 times larger than that with the G-parity BC \cite{
  RBC:2023ynh}, while the results for $\Im A_0$ and $\Im A_2$ are
consistent with each other as you can see in Table \ref{tab:xi0}
\subref{subtab:latt-ImA0} and \subref{subtab:latt-ImA2}.
Here we use the results for $\Im A_0$ and $\Im A_2$ obtained using the
G-parity BC to evaluate $|\epsK|$ because their errors are much less
than those with the periodic BC.

One remaining caveat is that the $\xi_0$ and $\xi_2$ in Ref.~\cite{
  RBC:2020kdj, RBC:2023ynh} are calculated in the isospin symmetric
limit (\emph{i.e.} $m_u = m_d$).
The isospin breaking effects on $\eps'/\eps$ are studied in
Refs.~\cite{ Gardner:1999wb, Cirigliano:2003nn}.
These studies conclude that the isospin violation correction in the CP
violation correction for $\eps'$ is below 15\% within the
uncertainties of large $N_c$ estimates for the low energy constants.
Since $\xi_0$ has an effect of about $-7.8\%$ on $\epsK$, the maximum of the
isospin violation effect, 15\% of $\xi_0$ amounts to $\pm 1\%$
correction for $\epsK$.
Here we neglect this effect of isospin breaking without loss of
generality in our conclusion\footnote{ Note that the gap $\Delta \epsK$
between the experimental value and theoretical expectation value of
$|\epsK|$ is about 35\%. Hence, the 1\% effect of isospin breaking
cannot change the conclusion of this paper. For more details, refer to
Sec.~\ref{sec:conclude}.}.
%

%
%\wlee{\\ EDIT by wlee \\}
%

%% file: tab_xi0.tex
%
% Table of input parameters
%
\begin{table}[htbp]
  \caption{ Input parameter $\xi_0$. }
  \label{tab:xi0}
  \renewcommand{\arraystretch}{1.4}
  \subtable[Experimental results for $\Re A_0$, $\Re A_2$ and etc.]{
    \begin{ruledtabular}
      \begin{tabular}{ l l l }
        parameter & value & Ref. \\ \hline
        $\Re A_0$ & $3.3201(18) \times 10^{-7} \GeV$ &
        \cite{ Blum:2015ywa, Bai:2015nea}
        \\
        $\Re A_2$ & $1.4787(31) \times 10^{-8} \GeV$ &
        \cite{ Blum:2015ywa}
        \\ 
        $\omega$  & $0.04454(12)$ &
        \cite{ Blum:2015ywa}
        \\ 
        $|\epsK|$ & $2.228(11) \times 10^{-3}$ &
        \cite{ Workman:2022ynf}
        \\
        $\Re(\eps'/\eps)$ & $1.66(23) \times 10^{-3}$ &
        \cite{ Workman:2022ynf}
      \end{tabular}
    \end{ruledtabular}
    \label{subtab:exp-ReA0-ReA2}
  } %%% \subtable
  \subtable[Lattice QCD results for $\Im A_0$ in units of $1.0 \times
    10^{-11} \GeV$.]{
    \begin{ruledtabular}
      \begin{tabular}{ l l l l }
        parameter & method & value & Ref. \\ \hline
        $\Im A_0$ & G-parity BC & \red{$-6.98(62)(144)$} &
        \cite{ RBC:2020kdj} \\
        $\Im A_0$ & periodic BC & \blue{$-8.7(12)(26)$}   &
        \cite{ RBC:2023ynh}
      \end{tabular}
    \end{ruledtabular}
    \label{subtab:latt-ImA0}
  } %%% \subtable
  \subtable[Lattice QCD results for $\Im A_2$ in units of $1.0 \times
    10^{-13} \GeV$.]{
    \begin{ruledtabular}
      \begin{tabular}{ l l l l }
        parameter & method & value & Ref. \\ \hline
        $\Im A_2$ & G-parity BC & \red{$-8.34(103)$}  &
        \cite{ RBC:2020kdj}  \\
        $\Im A_2$ & periodic BC & \blue{$-5.91(13)(175)$}  &
        \cite{ RBC:2023ynh}
      \end{tabular}
    \end{ruledtabular}
    \label{subtab:latt-ImA2}
  } %%% \subtable
  \subtable[$\xi_0$ in units of $1.0 \times 10^{-4}$.]{
    \begin{ruledtabular}
      \begin{tabular}{lllc}
        parameter & method & value & Ref. \\ \hline
        $\xi_0$ & indirect & \red{ $-1.738(177)$ }
        & \cite{ RBC:2020kdj} 
        \\
        $\xi_0$ & direct  & \red{ $-2.102(472)$ }
        & \cite{ RBC:2020kdj}
      \end{tabular}
    \end{ruledtabular}
    \label{subtab:xi0}
  }
\end{table}

%% file: ip_xi_LD.tex
%
% ip_xi_LD.tex
%
\subsection{$\xi_\text{LD}$}
\label{ssec:xi_LD}
The dispersive long distance (LD) effect on $\epsK$ are explained
in Eqs.~\eqref{eq:xiLD} and \eqref{eq:mLD} in Sec.~\ref{sssec:LD}.

Lattice QCD tools to calculate the dispersive LD effect,
$\xi_\text{LD}$ are well established in Ref.~\cite{ Christ:2012se,
  Bai:2014cva, Christ:2015pwa}.
Recently, there have been a number of attempts to calculate
$\xi_\text{LD}$ on the lattice \cite{ Christ:2015phf, Bai:2016gzv}.
These attempts in Refs.~\cite{ Christ:2015phf, Bai:2016gzv} belong to
a category of exploratory study rather than that of precision
measurement.

The net contribution of $\xi_\text{LD}$ to $\epsK$ in
Eqs.~\eqref{eq:xiLD} and \eqref{eq:mLD} turns out to be of the same
order of magnitude as $\xi_0$ using chiral perturbation theory
\cite{Buras2010:PhysLettB.688.309}.
They provide the formula for $\xi_\text{LD}$ as follows,
\begin{align}
  \xi_\text{LD} &= \black{-0.4(3) \times \frac{\xi_0}{\sqrt{2}} }\,,
  \label{eq:xiLD:BGI}
\end{align}
Here we use the results of the indirect method for $\xi_0$ in Table
\ref{tab:xi0}.
We call this method the BGI estimate for $\xi_\text{LD}$.
This also indicates that $\xi_\text{LD}$ is around a $5(4)\% \sim
6(5)\%$ correction to $\epsK$.
This claim is highly consistent with the estimate of about $2\%$ in
Ref.~\cite{ Christ:2012se, Christ:2014qwa}:
\begin{align}
  \xi_\text{LD} = (0 \pm 1.6)\%\,.
  \label{eq:xiLD:christ}
\end{align}
We call this method the RBC-UKQCD estimate for $\xi_\text{LD}$.
In this paper, we use both method to estimate $\xi_\text{LD}$.
%

%
%\wlee{\\ EDIT by wlee \\}
%

%% file: ip_m_t.tex
%
% ip_m_t.tex
%
\subsection{Top quark mass}
\label{ssec:top}
Results for the pole mass ($M_t$) of top quarks are summarized in
Table \ref{tab:m_t}.
The pole mass ($M_t$) and the $\MSb$ mass ($m_t(\mu)$)
renormalized at scale $\mu$ are related as follows,
\begin{align}
  \frac{m_t(\mu)}{M_t} &= z(\mu) = \frac{ Z_\text{OS} }{ Z_{\MSb}} \,.
\end{align}
Here, $Z_\text{OS}$ is the renormalization factor in the on-shell
scheme, and $Z_{\MSb}$ is the renormalization factor in the $\MSb$
scheme.
In Table \ref{tab:m_t}, $m_t(m_t)$ is the top scale-invariant quark
mass at the scale set to $\mu = m_t(\mu)$.
Technical details on how to obtain the scale-invariant top quark mass
$m_t(m_t)$ in the $\MSb$ scheme at the four-loop level is explained in
Ref.~\cite{ Bailey:2018feb}.
In Table \ref{tab:m_t}, the first error of $m_t(m_t)$ comes from
the error of the pole mass $M_t$ and the second error comes from
truncation of the higher loops in the conversion formula.
Technical details on this are described in Ref.~\cite{
  Bailey:2018feb}.
We have neglected the renormalon ambiguity and corrections due to the
three-loop fermion mass such as $m_b$ (bottom quark mass) and $m_c$
(charm quark mass).
Our results for $m_t(m_t)$ are consistent with those of RUNDEC
\cite{Chetyrkin2000:RunDec}.

\input{tab_m_t.tex}

In Fig. \ref{fig:m_t}, we present time evolution of the top quark pole
mass $M_t$.
We find that the average value shifts downward by 0.47\% over time and
the error shrinks fast down to 30\% of the beginning error (2012)
thanks to accumulation of high statistics in the LHC experiments.
\input{fig_m_t.tex}
%

%
%\wlee{\\ EDIT by wlee \\}
%

%% file: tab_M_t.tex
\begin{table}[h]
%%  \footnotesize
  \renewcommand{\arraystretch}{1.3}
  \caption{Results for $M_t$ in units of $\GeV$.}
  \label{tab:m_t}
  \begin{ruledtabular}
    \begin{tabular}{ l l l l }
      {\small Collaboration} & $M_t$ & $m_t(m_t)$ & Ref.
      \\ \hline
      PDG 2021 & 172.76(30) & 162.96(28)(17) & \cite{Zyla:2020zbs}
      \\
      PDG 2022 & 172.69(30) & 162.90(28)(17) & \cite{Workman:2022ynf}
      \\
      PDG 2023 & 172.69(30) & 162.90(28)(17) & \cite{Workman:2022ynf}
      \\
      PDG 2024 & 172.57(29) & \red{ 162.77(27)(17) } & \cite{ ParticleDataGroup:2024cfk}%p1379
    \end{tabular}
  \end{ruledtabular}
\end{table}

%% file: fig_m_t.tex
\begin{figure}[h!]
  \includegraphics[width=1.0\linewidth]{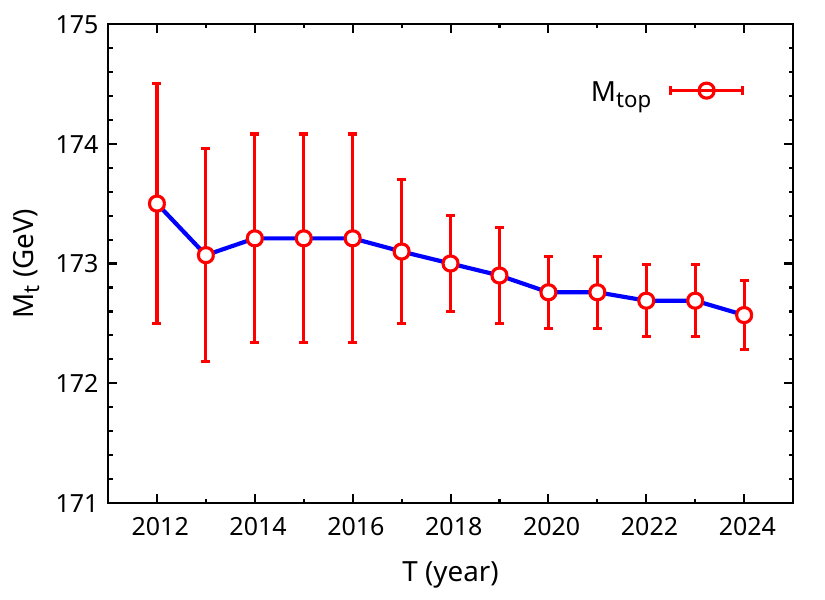}
  \caption{History of $M_t$ (top quark pole mass).}
  \label{fig:m_t}
\end{figure}

%% file: ip_m_c.tex
%
% ip_m_c.tex
%
\subsection{Charm quark mass }
\label{ssec:m_c}
In Table \ref{tab:m_c} we present lattice QCD results for the charm
quark mass $m_c(m_c)$ determined from lattice QCD (FLAG 2024) \cite{
  FlavourLatticeAveragingGroupFLAG:2024oxs}.
We find that the results for $m_c(m_c)$ with $N_f=2+1$ in Table
\ref{tab:m_c} \subref{stab:m_c-2+1} are consistent with one another.
The results for $m_c(m_c)$ with $N_f=2+1+1$ are inconsistent between
ETM and the rest.
This ends up with a very large error in the combined results for
$m_c(m_c)$ with $N_f=2+1+1$ in Table \ref{tab:m_c}
\subref{stab:m_c-sum}.
We suspect that there might be some issues unresolved in the results
of ETM mainly because the rest are consistent with the results with
$N_f = 2+1$.
Hence, we prefer the results for $m_c (m_c)$ with $N_f = 2+1$ to
evaluate $\epsK$.

\input{tab_m_c.tex}
%

%
%\wlee{\\ EDIT by wlee \\}
%

%% file: tab_m_c.tex
\begin{table}[h!]
%%  \footnotesize
  \renewcommand{\arraystretch}{1.3}
  \caption{Results for $m_c(m_c)$ in units of $\GeV$.}
  \label{tab:m_c}
  \subtable[Summary of $m_c(m_c)$]{
    \begin{ruledtabular}
      \begin{tabular}{ l l l l }
        Collaboration   & $N_f$   & $m_c(m_c)$ & Ref.
        \\ \hline
        FLAG 2024       & $2+1$   & \red{1.278(6)}
        & \cite{ FlavourLatticeAveragingGroupFLAG:2024oxs}
        \\
        FLAG 2024       & $2+1+1$ & \blue{1.280(13)}
        & \cite{ FlavourLatticeAveragingGroupFLAG:2024oxs}
      \end{tabular}
    \end{ruledtabular}
    \label{stab:m_c-sum}
  } % subtable
  \vfill
  \subtable[$m_c(m_c)$ with $N_f=2+1$]{
    \begin{ruledtabular}
      \begin{tabular}{ l l l }
        Collaboration   & $m_c(m_c)$ & Ref.
        \\ \hline
        ALPHA 23       & 1.296(15)    & \cite{ Bussone:2023kag}
        \\
        ALPHA 21       & 1.296(19)    & \cite{ Heitger:2021apz}
        \\
        Petreczky 19   & 1.265(10)    & \cite{ Petreczky:2019ozv}
        \\
        JLQCD 16       & 1.2871(123)  & \cite{ Nakayama:2016atf}
        \\
        $\chi$QCD 14   & 1.304(5)(20) & \cite{ Yang:2014sea}
        \\
        HPQCD 10       & 1.273(6)     & \cite{ McNeile:2010ji}
        \\ \hline
        RBC/UKQCD 24\footnote{This is not included in the FLAG 2024 results for $N_f = 2+1$.}
                       & 1.292(12)    & \cite{ DelDebbio:2024hca}
      \end{tabular}
    \end{ruledtabular}
    \label{stab:m_c-2+1}
  } % subtable
  \vfill
  \subtable[$m_c(m_c)$ with $N_f=2+1+1$]{
    \begin{ruledtabular}
      \begin{tabular}{ l l l }
        Collaboration   & $m_c(m_c)$ & Ref.
        \\ \hline
        ETM 21A             & 1.339(22)$(^{+19}_{-10})$(10)
        & \cite{ExtendedTwistedMass:2021gbo}
        \\
        HPQCD 20A           & 1.2719(78)
        & \cite{Hatton:2020qhk}
        \\
        FNAL/MILC/TUMQCD 18 & 1.273(4)(1)(10)
        & \cite{FermilabLattice:2018est}
        \\
        HPQCD 14A           & 1.2715(95)
        & \cite{Chakraborty:2014aca}
        \\
        ETM 14A             & 1.3478(27)(195)
        & \cite{Alexandrou:2014sha}
        \\
        ETM 14              & 1.348(46)
        & \cite{EuropeanTwistedMass:2014osg}
      \end{tabular}
    \end{ruledtabular}
    \label{stab:m_c-2+1+1}
  } % subtable
\end{table}

%% file: ip_W_mass.tex
%
% ip_W_mass.tex
%
\subsection{$W$-boson mass}
\label{ssec:m_W}
In Fig.~\ref{fig:m_W}, we plot the $W$ boson mass ($M_W$) as a
function of time.
The corresponding values for $M_W$ are presented in Table
\ref{tab:m_W}.
\input{fig_m_W.tex}
\input{tab_m_W.tex}

In Fig.~\ref{fig:m_W}, the light-green band represents the standard
model (SM) prediction, the red circles represent the PDG results from
the experimental summary, and the brown cross represents the CDF-2022
result.
Here we find that all the results of PDG are consistent with the
SM prediction, while the CDF result deviates from the SM
prediction by $\approx 7\sigma$. 

We use the SM-2023 result for $M_W$ to evaluate $\epsK$.

%
%\wlee{\\ EDIT by wlee \\}
%

%% file: fig_m_W.tex
\begin{figure}[h!]
  %%%\vspace{-3mm}
  \includegraphics[width=1.0\linewidth]{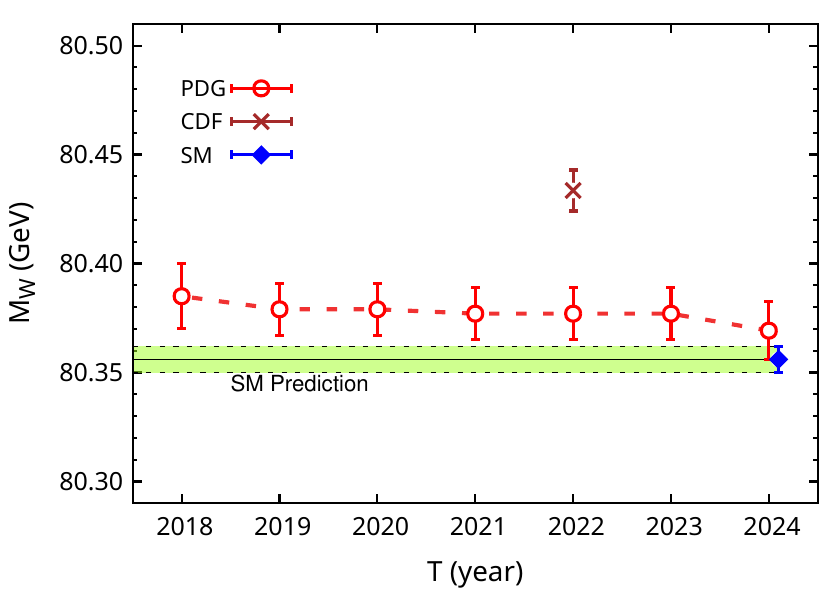}
  \caption{Time evolution of $M_W$.}
  \label{fig:m_W}
\end{figure}

%% file: tab_m_W.tex
\begin{table}[h!]
  %%%\vspace{-3mm}
  \renewcommand{\arraystretch}{1.35}
  \caption{ Values of $W$-boson mass $M_W$ in Fig.~\ref{fig:m_W}. }
  \label{tab:m_W}
  \begin{ruledtabular}
    \begin{tabular}{ l l l }
      Source & $M_W$ (GeV) & Ref.
      \\ \hline
      %SM-2024  & \red{ 80.356(6) }   & \cite{ Workman:2022ynf}   \\
      SM-2024  & \red{ 80.353(6) }   & \cite{ ParticleDataGroup:2024cfk}    \\%p815,s54p3
      CDF-2022 & \blue{80.4335(94)}  & \cite{ CDF:2022hxs}       \\
      PDG-2024 &       80.3692(133)  & \cite{ ParticleDataGroup:2024cfk} \\
      PDG-2023 &       80.377(12)    & \cite{ Workman:2022ynf}   \\
      PDG-2022 &       80.377(12)    & \cite{ Workman:2022ynf}   \\
      PDG-2021 &       80.377(12)    & \cite{ Zyla:2020zbs}      \\
      PDG-2020 &       80.379(12)    & \cite{ Zyla:2020zbs}      \\
      PDG-2019 &       80.379(12)    & \cite{ Tanabashi:2018oca} \\
      PDG-2018 &       80.385(15)    & \cite{ Patrignani:2016xqp}
    \end{tabular}
  \end{ruledtabular}
\end{table}

%% file: ip_eta.tex
%
% ip_eta.tex
%
\subsection{Higher order QCD corrections}
\label{ssec:eta_i}
We summarize higher order QCD corrections $\eta_i$ in Table
\ref{tab:input-eta}.
There are two sets of $\eta_i$: one is $\eta_i$ of $c-t$ unitarity
(the \texttt{traditional method} used in Ref.~\cite{ Bailey:2015tba,
  Bailey:2018feb}, and the other is $\eta_i$ ($=\eta_i^\text{BGS}$) of
$u-t$ unitarity (the \texttt{BGS method} used in Ref.~\cite{
  Brod:2019rzc}).
The traditional method for $\eta_i$ of $c-t$ unitarity is described in
Subsection \ref{ssec:master-tr-ctu} (Eqs.~\eqref{eq:epsK_SM_0} and
\eqref{eq:X_SD}), and the BGS method for $\eta_i$ of $u-t$ unitarity
is explained in Subsection \ref{ssec:BGS}.
We present $\eta_i$ of $c-t$ unitarity (the traditional method) in
Table \ref{tab:input-eta} \subref{tab:eta_ctu}, and $\eta_i$ of $u-t$
unitarity (the BGS method) in Table \ref{tab:input-eta}
\subref{tab:eta_utu}.

\input{tab_eta}

The BGS method ($\eta_{ut}^\text{BGS}$) are supposed to have better
convergence with respect to the charm quark mass contribution
\cite{Brod:2019rzc}.
%

%
%\wlee{\\ EDIT by wlee \\}
%

%% file: tab_eta.tex
\begin{table}[h]
  \renewcommand{\arraystretch}{1.2}
  \caption{Higher order QCD corrections.}
  \label{tab:input-eta}
	\subtable[$\eta_{i}$ of $c-t$ unitarity (traditional method)]{
	 \begin{ruledtabular}
    	  	\begin{tabular}[b]{@{\quad} c @{\qquad}l @{\qquad}c @{\quad}}
      	  	Input & Value & Ref. \\ \hline
      		$\eta_{cc}$ & $1.72(27)$   & \cite{Bailey:2015tba} 
		\\
		$\eta_{tt}$ & $0.5765(65)$ & \cite{Buras2008:PhysRevD.78.033005} 
		\\	
		$\eta_{ct}$ & $0.496(47)$  & \cite{Brod2010:prd.82.094026}
		\end{tabular}
  	\end{ruledtabular}
	\label{tab:eta_ctu}
	} % subtable
	\vfill
	\subtable[$\eta_{i}$ of $u-t$ unitarity (BGS method)]{
	 \begin{ruledtabular}
   		\begin{tabular}[b]{@{\quad} c @{\qquad}l @{\qquad}c @{\quad}}
	        Input & Value & Ref.
        	\\ \hline
	        $\eta_{tt}^{\text{BGS}}$ & $0.55(1\pm4.2\%\pm0.1\%)$   & \cite{Brod:2019rzc}
	        \\
	        $\eta_{ut}^{\text{BGS}}$ & $0.402(1\pm1.3\%\pm0.2\%\pm0.2\%)$ & \cite{Brod:2019rzc}
	      \end{tabular}
	  \end{ruledtabular}
	  \label{tab:eta_utu}
	  } %subtable
\end{table}

%% file: ip_other.tex
%
% ip_other.tex
%
\subsection{Other Input Parameters }
\label{ssec:other}
Other input parameters are summarized in Table \ref{tab:other}.
They are the same as Ref.~\cite{ Bailey:2018feb} except for
the Fermi constant $G_F$, and the kaon decay constant $F_K$.
For the Fermi constant, we use the updated results of PDG-2024
\cite{ ParticleDataGroup:2024cfk}.
For the kaon decay constant, we use the lattice QCD results of
FLAG-2024 \cite{ FlavourLatticeAveragingGroupFLAG:2024oxs}.

\input{tab_other}

%
%\wlee{\\ EDIT by wlee \\}
%

%% file: tab_other.tex
\begin{table}[h]
  \caption{Other input parameters.}
  \label{tab:other}
  \renewcommand{\arraystretch}{1.2}
  \begin{ruledtabular}
    \begin{tabular}{@{\quad} c @{\qquad} l @{\qquad} c @{\quad}}
      Input & Value & Ref. \\ \hline
      $G_{F}$
      & \red{ $1.1663788(6) \times 10^{-5} \; \GeV^{-2}$ }
      &\cite{ ParticleDataGroup:2024cfk} \\ \hline
      $\theta$
      & $43.52(5)^{\circ}$
      &\cite{ ParticleDataGroup:2024cfk} \\ \hline
      $m_{K^{0}}$
      & $497.611(13) \;\MeV$
      &\cite{ ParticleDataGroup:2024cfk} \\ \hline
      $\Delta M_{K}$
      & $3.484(6) \times 10^{-12} \;\MeV$
      &\cite{ ParticleDataGroup:2024cfk} \\ \hline
      $F_K$
      & \red{ $155.7(3) \;\MeV$ }
      &\cite{ FlavourLatticeAveragingGroupFLAG:2024oxs}
    \end{tabular}
  \end{ruledtabular}
\end{table}

%% file: result.tex
%
% result.tex
%
\section{Results} 
\label{sec:result}
Here we present results for $|\epsK|$ obtained with various combination
of input parameters.
We report results for $|\epsK|^\text{SM}$ obtained using the
\texttt{traditional} method with $\eta_i$ of $c-t$ unitarity in
Subsection \ref{ssec:c-t-uni}, and those obtained using the
\texttt{BGS} method with $\eta_i$ of $u-t$ unitarity in Subsection
\ref{ssec:u-t-uni}.

\subsection{$\eta_i$ of $c-t$ unitarity (the traditional method)}
\label{ssec:c-t-uni}
Here we present results for $|\epsK|^\text{SM}$ obtained using the
\texttt{traditional} method with $\eta_i$ of $c-t$ unitarity.
We classify them into two sub-categories: (1) results for
$|\epsK|^\text{SM}$ obtained using the \texttt{RBC-UKQCD} estimate for
$\xi_\text{LD}$, and (2) those obtained using the \texttt{BGI}
estimate for $\xi_\text{LD}$.
The former are presented in Subsubsection \ref{sssec:ctu:rbc} and
the latter in Subsubsection \ref{sssec:ctu:bgi}.

\subsubsection{RBC-UKQCD estimate for $\xi_\text{LD}$}
\label{sssec:ctu:rbc}
In Table \ref{tab:epsK:ctu:rbc:indirect}, we present results for
$|\epsK|^\text{SM}$ and $\Delta\epsK$ obtained using the
\texttt{indirect} method for $\xi_0$, the \texttt{RBC-UKQCD} estimate
for $\xi_\mathrm{LD}$, the \texttt{traditional} method for $\eta_i$ of
$c-t$ unitarity, the FLAG-24 results for $\BK$, and the AOF results
for Wolfenstein parameters.
Here the superscript ${}^\text{SM}$ represents the theoretical
expectation value of $|\epsK|$ obtained directly from the SM, and
$\Delta \epsK \equiv |\epsK|^\text{Exp} - |\epsK|^\text{SM}$.
The superscript ${}^\text{Exp}$ represents the experimental result for
$|\epsK|$.
In Table \ref{tab:epsK:ctu:rbc:indirect}, we find that the theoretical
results for $|\epsK|^\text{SM}$ obtained with lattice QCD inputs
including exclusive $\Vcb$ have $5.1\sigma \sim 4.1\sigma$ tension
with the experimental results for $|\epsK|^\text{Exp}$.
In contrast, we also find that there is no tension with those obtained
using inclusive $\Vcb$, which is obtained using the heavy quark
expansion and QCD sum rules.
%
\input{tab_epsK_ctu_rbc_indir.tex}
In Table \ref{tab:epsK:ctu:rbc:direct}, we report results for
$|\epsK|^\text{SM}$ and $\Delta \epsK$ evaluated using the
\texttt{direct} method for $\xi_0$, the \text{RBC-UKQCD} estimate for
$\xi_\text{LD}$, the \texttt{traditional} method for $\eta_i$ of $c-t$
unitarity, the FLAG-24 results for $\BK$, and the AOF results for
Wolfenstein parameters.
In Table \ref{tab:epsK:ctu:rbc:direct}, we find that the theoretical
expectation values for $|\epsK|^\text{SM}$ determined from lattice QCD
inputs including exclusive $\Vcb$ have $5.2\sigma \sim 4.1\sigma$
tension with the experimental results for $|\epsK|^\text{Exp}$.
This is consistent with what we observe in Table
\ref{tab:epsK:ctu:rbc:indirect}.
%
\input{tab_epsK_ctu_rbc_dir.tex}
In Fig.~\ref{fig:epsK:cmp:ctu:rbc:indirect}, we make a plot of results
for $\epsK$ calculated directly from the SM with the lattice QCD
inputs described in Section \ref{sec:input}.
In Fig.~\ref{fig:epsK:cmp:ctu:rbc:indirect}
\subref{fig:epsK-ex:ctu:rbc:indirect}, the blue curve which encircles
the histogram represents the theoretical evaluation of $\epsK$ using
the FLAG-24 $\BK$, the AOF for Wolfenstein parameters, the exclusive
$\Vcb$ (FNAL/MILC-22 in Table \ref{tab:Vcb} \subref{tab:ex-Vcb}), the
indirect method for $\xi_0$, the RBC-UKQCD estimate for
$\xi_\text{LD}$, and the traditional method for $\eta_i$ of $c-t$
unitarity.
The red curves in Fig.~\ref{fig:epsK:cmp:ctu:rbc:indirect} represent
the experimental results for $|\epsK|$ ($|\epsK|^\text{Exp}$).
In Fig.~\ref{fig:epsK:cmp:ctu:rbc:indirect}
\subref{fig:epsK-in:ctu:rbc:indirect}, the blue curve represents the
same as in Fig.~\ref{fig:epsK:cmp:ctu:rbc:indirect}
\subref{fig:epsK-ex:ctu:rbc:indirect} except for using the inclusive
$\Vcb$ (1S-23 in Table \ref{tab:Vcb} \subref{tab:in-Vcb}).
%
%
\input{ fig_epsK_in_ex_ctu_rbc.tex}

In Fig.~\ref{fig:epsK:cmp:ctu:rbc:indirect}, the results for $|\epsK|$
are presented in units of $1.0\times 10^{-3}$ as in Table
\ref{tab:epsK:ctu:rbc:indirect}.
From Fig.~\ref{fig:epsK:cmp:ctu:rbc:indirect} and Table
\ref{tab:epsK:ctu:rbc:indirect}, we find that the theoretical
expectation value of $|\epsK|^\text{SM}$ with lattice QCD inputs (with
exclusive $\Vcb$ of FNAL/MILC-22) has $5.10\sigma$ tension with the
experimental result $|\epsK|^\text{Exp}$, while there is no tension in
the inclusive $\Vcb$ channel (heavy quark expansion based on the OPE
and QCD sum rules).

Here we use the results for exclusive $\Vcb$ from FNAL/MILC-22 as a
representative sample intentionally, since it contains the most
comprehensive analysis of the $\BtoDst$ decays at both zero recoil and
non-zero recoil, and it incorporate experimental results from both
BELLE and BABAR.
Another reason is that it comes from a single decay channel of
$\BtoDst$ and so the error estimation is relatively clean compared
with other results such as FLAG-24, HFLAV-23, and HPQCD-23, because
one does not need to worry about caveats due to the unwanted
correlation between the semileptonic form factors from lattice QCD by
sharing the same gauge configurations in common for the measurement.
In Fig.~\ref{fig:depsK:sum:ctu:rbc:his}
\subref{fig:depsK:ctu:rbc:his}, we plot the $\Delta\epsK \equiv
|\epsK|^\text{Exp} - |\epsK|^\text{SM}$ in units of $\sigma$ (which is
the total error of $\Delta\epsK$) as a function of the time.
Here we use the same $|\epsK|^\text{SM}$ as in
Fig.~\ref{fig:epsK:cmp:ctu:rbc:indirect}
\subref{fig:epsK-ex:ctu:rbc:indirect}.
We began to monitor $\Delta\epsK$ in 2012 when several lattice QCD
results for $\BK$ obtained using different discretization methods
for the light and strange quarks became consistent with one another
within statistical uncertainty.
In 2012, $\Delta\epsK$ was $2.5\sigma$, but now it is $5.1\sigma$.
In order to understand the change of $\Delta\epsK/\sigma$ with respect
to time, we perform an additional analysis on the average and error.
%
\input{ fig_epsK_history_ctu_rbc.tex}

In Fig.~\ref{fig:depsK:sum:ctu:rbc:his}
\subref{fig:depsK+sigma:ctu:rbc:his}, we plot the average
$\Delta\epsK$ and the error $\sigma_{\Delta\epsK}$ as a function of
time.
Here we find that the average of $\Delta\epsK$ has increased with some
fluctuations by $\cong 49\%$ during the period of 2012--2024, and
its error $\sigma_{\Delta\epsK}$ has decreased with some ripples by
$\cong 27\%$ during the same period.
These two effects interfere constructively to produce the $5.1\sigma$
tension in $\Delta\epsK$ in 2024.
The decrease in $\sigma_{\Delta\epsK}$ with respect to time comes from
significant progress in achieving higher precision on input parameters
during this period (2023--2024).
As time goes on, the lattice QCD calculations become more precise, and
the experimental results also get more accurate, which constructively
leads to the reduction in $\sigma_{\Delta\epsK}$.

In Table \ref{tab:err-budget:ctu:rbc}, we present the error budget for
$|\epsK|^\text{SM}$ evaluated using the AOF results of Wolfenstein
parameters, the FNAL/MILC-22 results for exclusive $\Vcb$, the FLAG-24
results for $\BK$, the RBC-UKQCD estimate for $\xi_\text{LD}$, the
indirect method for $\xi_0$, and the traditional method for $\eta_i$
of $c-t$ unitarity.
Here we find that the dominant error ($\cong 52\%$) in
$|\epsK|^\text{SM}$ comes from exclusive $\Vcb$, while the errors
coming from $\eta_{ct}$, $\bar{\eta}$, and $\eta_{cc}$ are
sub-dominant.
Hence, if we are to observe a gap $\Delta\epsK$ much greater than
$5.1\sigma$, it is essential to reduce the error in exclusive $\Vcb$
significantly.
%
\input{ tab_err_bud_ctu_rbc.tex}
In exclusive $\Vcb$, the error of $\Vcb$ comes from the two sources:
one is experimental and the other is theoretical.
The errors coming from the experiments are beyond the scope of this
paper, and so we focus on the errors coming from the theory (lattice
QCD).
The dominant error in the theoretical calculation of the semileptonic
form factors using lattice QCD tools comes from the heavy quark
discretization error (HQDE) for the charm quark in lattice QCD.
If one use the Fermilab action \cite{ El-Khadra:1996wdx}, the HQDE is
about 1.0\%, which is significantly larger than any other errors in
the theoretical side.
The Oktay-Kronfeld (OK) action \cite{ Oktay:2008ex} is improved up to
the $\lambda^3$ order at the tree level, while the Fermilab action is
improved up to the $\sim \lambda^1$ order.
In order to reduce the HQDE by a factor of $\lambda \sim 1/5$, there
are on-going efforts to use the OK action and improved currents \cite{
  Bailey:2020uon} to calculate the $B \to D^{(*)}$ semileptonic form
factors as in Ref.~\cite{ Bhattacharya:2023llf}.
\subsubsection{BGI estimate for $\xi_\text{LD}$}
\label{sssec:ctu:bgi}
Here, we present results for $|\epsK|^\text{SM}$ obtained using the
\texttt{BGI} estimate in Eq.~\eqref{eq:xiLD:BGI} for $\xi_\text{LD}$,
and the \texttt{traditional} method for $\eta_i$ of $c-t$ unitarity.

In Table \ref{tab:epsK:ctu:bgi:indirect}, we summarize results for
$|\epsK|^\text{SM}$ obtained using the \texttt{indirect} method for
$\xi_0$, the \texttt{BGI} estimate for $\xi_\text{LD}$, the
\texttt{traditional} method for $\eta_i$ of $c-t$ unitarity, the
FLAG-24 results for $\BK$, and the AOF results for Wolfenstein
parameters.
Here we find that the theoretical results for $|\epsK|^\text{SM}$
obtained using lattice QCD inputs including exclusive $\Vcb$ have
$4.7\sigma \sim 3.7 \sigma$ tension with the experimental results
for $|\epsK|^\text{Exp}$, while there is no tension with those
obtained using inclusive $\Vcb$.
%
\input{tab_epsK_ctu_bgi_indir.tex}
In Table \ref{tab:epsK:ctu:bgi:direct}, we report results for
$|\epsK|^\text{SM}$ evaluated using the \texttt{direct} method
for $\xi_0$ with the rest of the input parameters set to
the same as in Table \ref{tab:epsK:ctu:bgi:indirect}.
Here we find that the theoretical values for $|\epsK|^\text{SM}$
determined using lattice QCD inputs with exclusive $\Vcb$ have
$4.6\sigma \sim 3.7\sigma$ tension with the experimental results for
$|\epsK|^\text{Exp}$.
This is consistent with what we find from Table
\ref{tab:epsK:ctu:bgi:indirect}.
%
\input{tab_epsK_ctu_bgi_dir.tex}
In Fig.~\ref{fig:epsK:cmp:ctu:bgi} \subref{fig:epsK-ex:ctu:bgi}, the
blue curve represents the theoretical evaluation of $|\epsK|$ directly
from the standard model (SM) using the same input parameters as in
Fig.~\ref{fig:epsK:cmp:ctu:rbc:indirect}
\subref{fig:epsK-ex:ctu:rbc:indirect} except for the BGI estimate in
Eq.~\eqref{eq:xiLD:BGI} for $\xi_\text{LD}$.
The red curve in Fig.~\ref{fig:epsK:cmp:ctu:bgi} represents the
experimental results for $|\epsK|^\text{Exp}$.
In Fig.~\ref{fig:epsK:cmp:ctu:bgi} \subref{fig:epsK-in:ctu:bgi}, the
blue curve are the same as in Fig.~\ref{fig:epsK:cmp:ctu:bgi}
\subref{fig:epsK-ex:ctu:bgi} except for using inclusive $\Vcb$ (1S-23
in Table \ref{tab:Vcb} \subref{tab:in-Vcb}).
%
\input{fig_epsK_in_ex_ctu_bgi.tex}

In Fig.~\ref{fig:depsK:sum:ctu:bgi:his}
\subref{fig:depsK:ctu:bgi:his}, we plot $\Delta\epsK$ in units of
$\sigma$ (= the error of $\Delta\epsK$) as a function of time.
In 2012, $\Delta\epsK$ was $2.3\sigma$, but now it is $4.7\sigma$.
In order to understand this transition, we have done additional
analysis on the average and error.
%
%
\input{fig_epsK_history_ctu_bgi.tex}
In Fig.~\ref{fig:depsK:sum:ctu:bgi:his}
\subref{fig:depsK+sigma:ctu:bgi:his}, we plot the time evolution of
the average and error for $\Delta\epsK$.
Here we find that the average of $\Delta\epsK$ has increased by $\cong
53\%$ with some fluctuations during the period of 2012--2024, and its
error has decreased by $\cong 26\%$ with tiny ripples in the same
period.
These two effect has produced constructively the $4.7\sigma$ tension
in $\Delta\epsK$.
In Table \ref{tab:err-budget:ctu:bgi}, we present the error budget for
$|\epsK|^\text{SM}$ in Fig.~\ref{fig:epsK:cmp:ctu:bgi}
\subref{fig:epsK-ex:ctu:bgi}.
Here, we find that the dominant error ($\cong 50\%$) in
$|\epsK|^\text{SM}$ comes from exclusive $\Vcb$.
Hence, if we are to observe the gap $\Delta\epsK$ much larger than
$5.0\sigma$, it is essential to reduce the error in $\Vcb$
significantly.
%
\input{ tab_err_bud_ctu_bgi.tex}
\subsection{$\eta_i$ of $u-t$ unitarity (the BGS method)}
\label{ssec:u-t-uni}
Here we present results for $|\epsK|^\text{SM}$ obtained using the
BGS method with $\eta_i^\text{BGS}$ of $u-t$ unitarity.
We classify them into two sub-categories: (1) results for
$|\epsK|^\text{SM}$ obtained using the RBC-UKQCD estimate for
$\xi_\text{LD}$, and (2) those obtained using the BGI estimate
for $\xi_\text{LD}$.
The former are presented in Subsubsection \ref{sssec:utu:rbc}, and the
latter in Subsubsection \ref{sssec:utu:bgi}.

\subsubsection{RBC-UKQCD estimate for $\xi_\text{LD}$ (with BGS)}
\label{sssec:utu:rbc}
In Table \ref{tab:epsK:utu:rbc:indirect}, we present results for
$|\epsK|^\text{SM}$ and $\Delta \epsK$ obtained using the
\texttt{indirect} method for $\xi_0$, the \texttt{RBC-UKQCD} estimate
for $\xi_\text{LD}$, the \texttt{BGS} method for $\eta_i$
($=\eta_i^\text{BGS}$) of $u-t$ unitarity, the FLAG-24 results for
$\BK$, and the AOF for Wolfenstein parameters.
Here we find a central value mismatch (CVM) $\delta \epsK^\text{BGS}$
in values for $|\epsK|^\text{SM}$ between the traditional method for
$\eta_i$ of $c-t$ unitarity and the BGS method for $\eta_i$ of $u-t$
unitarity.
For example, the FNAL/MILC-22 results for $|\epsK|^\text{SM}$ in units
of $10^{-3}$ are
\begin{align}
  |\epsK|^\text{SM}_{c-t} &= 1.453 \pm 0.152  & &
  \text{ from Table \ref{tab:epsK:ctu:rbc:indirect} }
  \\
  |\epsK|^\text{SM}_{u-t} &= 1.484 \pm 0.137  & &
  \text{ from Table \ref{tab:epsK:utu:rbc:indirect} }
  \\
  \delta \epsK^\text{BGS} &\equiv
  |\epsK|^\text{SM}_{u-t} - |\epsK|^\text{SM}_{c-t}
  \\
  &= 0.032
\end{align}
where the subscript $c-t$ ($u-t$) represents the traditional method
for $\eta_i$ of $c-t$ unitarity (the BGS method for $\eta_i$ of
$u-t$ unitarity).

This CVM comes from a number of small and tiny approximations
introduced in the BGS method when they obtain a simplified form of
$\eta_i^\text{BGS}$ in Eqs.~\eqref{eq:eta:BGS:tt-1} and
\eqref{eq:eta:BGS:ut-1}.
We count this CVM $\delta \epsK^\text{BGS}$ as an additional error in
the BGS method.
Hence, the total error is
\begin{align}
  \sigma_t^\text{BGS} &=
  \sqrt{ [\sigma_1^\text{BGS}]^2 + [\delta \epsK^\text{BGS}]^2}
  \label{eq:BGS:sigma_t-1}
\end{align}
where $\sigma_t^\text{BGS}$ represents the total error, and
$\sigma_1^\text{BGS}$ represents the errors coming from the input
parameters.
In the above example of FNAL/MILC-22,
\begin{align}
  \sigma_t^\text{BGS} &= \sqrt{0.133^2 + 0.032^2}
  \cong 0.137
\end{align}
In Table \ref{tab:epsK:utu:rbc:indirect}, we present
$\sigma_1^\text{BGS}$, $\delta\epsK^\text{BGS}$, and
$\sigma_t^\text{BGS}$ to provide some sense on numerical size of them.
In those tables which appear later in the paper, we will provide
only the total errors $\sigma_t^\text{BGS}$.
From Table \ref{tab:epsK:utu:rbc:indirect}, we find that the
theoretical results for $|\epsK|^\text{SM}$ obtained using lattice QCD
inputs including exclusive $\Vcb$ and the BGS method ($u-t$ unitarity)
have $5.7\sigma \sim 4.2\sigma$ tension with the experimental result
$|\epsK|^\text{Exp}$.
We also find that the tension disappears for those obtained using
inclusive $\Vcb$, which is obtained using the heavy quark expansion
and QCD sum rules.
%
\input{ tab_epsK_utu_rbc_indir.tex}
In Table \ref{tab:epsK:utu:rbc:direct}, we report results for
$|\epsK|^\text{SM}$ and $\Delta \epsK$ determined using the
\texttt{direct} method for $\xi_0$, the \texttt{RBC-UKQCD} estimate
for $\xi_\text{LD}$, the \texttt{BGS} method for $\eta_i$ of $u-t$
unitarity. the FLAG-24 results for $\BK$, and the AOF for Wolfenstein
parameters.
Here the error of $|\epsK|^\text{SM}$ represents the total error
$\sigma_t^\text{BGS}$ in Eq.~\eqref{eq:BGS:sigma_t-1}.
From Table \ref{tab:epsK:utu:rbc:direct}, we find that the theoretical
determination of $|\epsK|^\text{SM}$ obtained using lattice QCD inputs
including exclusive $\Vcb$ has $5.7\sigma \sim 4.3\sigma$ tension with
the experimental results for $|\epsK|^\text{Exp}$.
This is quite consistent with what we observe in Table
\ref{tab:epsK:utu:rbc:indirect}.
%
\input{ tab_epsK_utu_rbc_dir.tex}
In Fig.~\ref{fig:epsK:cmp:ctu:utu:rbc}
\subref{fig:epsK-ex:ctu:rbc:traditional}, we present results for
$|\epsK|^\text{SM}$ obtained using the \texttt{traditional} method for
$\eta_i$ of $c-t$ unitarity.
Fig.~\ref{fig:epsK:cmp:ctu:utu:rbc} \subref{fig:epsK-ex:utu:rbc:BGS}
is the same kind of a plot except for using the \texttt{BGS} method
for $\eta_i$ of $u-t$ unitarity.
Here the blue curve which encircles the histogram represents the
theoretical determination of $|\epsK|$ using the FLAG-24 results for
$\BK$, the AOF for Wolfenstein parameters, the FNAL/MILC-22 results
for exclusive $\Vcb$, the indirect method for $\xi_0$, and the
RBC-UKQCD estimate for $\xi_\text{LD}$.
The red curves in Fig.~\ref{fig:epsK:cmp:ctu:utu:rbc} represent
the experimental results for $|\epsK|^\text{Exp}$.
%
\input{ fig_epsK_ex_ctu_utu_rbc.tex}

From Fig.~\ref{fig:epsK:cmp:ctu:utu:rbc} and Table
\ref{tab:epsK:utu:rbc:indirect}, we find that the theoretical results
for $|\epsK|^\text{SM}$ with lattice QCD inputs with exclusive $\Vcb$
of the FNAL/MILC-22 and the \texttt{BGS} method for $\eta_i$ of $u-t$
unitarity have $5.4 \sigma$ tension with the experimental result
$|\epsK|^\text{Exp}$, while there is no tension in the inclusive
$\Vcb$ channel.
Here we use the FNAL/MILC-22 results for exclusive $\Vcb$ as a
representative sample for the same reason as explained in
Subsubsection \ref{sssec:ctu:rbc}.
In Table \ref{tab:err-budget:utu:rbc}, we present the error budget for
$|\epsK|^\text{SM}$ obtained using the AOF of Wolfenstein parameters,
the FNAL/MILC-22 results for exclusive $\Vcb$, the FLAG-24 results for
$\BK$, the RBC-UKQCD estimate for $\xi_\text{LD}$, the indirect method
for $\xi_0$, and the \texttt{BGS} method for $\eta_i$ of $u-t$
unitarity.
Here we find that the dominant error ($\cong 63\%$ ) in
$|\epsK|^\text{SM}$ comes from exclusive $\Vcb$, while the errors due
to $\bar\eta$ and $\eta_{tt}^\text{BGS}$ are subdominant.
Hence, if we are to observe $\Delta\epsK$ much larger than
$5.4\sigma$, it is essential to reduce the error in exclusive $\Vcb$
significantly.
%
\input{ tab_err_bud_utu_rbc.tex}
\subsubsection{BGI estimate for $\xi_\text{LD}$ (with BGS)}
\label{sssec:utu:bgi}
Here we present results for $|\epsK|^\text{SM}$ obtained using the
\texttt{BGI} estimate for $\xi_\text{LD}$, and the \texttt{BGS} method
for $\eta_i$ of $u-t$ unitarity.

In Table \ref{tab:epsK:utu:bgi:indirect}, we present results for
$|\epsK|^\text{SM}$ obtained using the \texttt{indirect} method for
$\xi_0$, the \texttt{BGI} estimate for $\xi_\text{LD}$, the
\texttt{BGS} method for $\eta_i$ of $u-t$ unitarity, the FLAG-24
results for $\BK$, and the AOF results for Wolfenstein parameters.
Here we find that the theoretical expectation values of
$|\epsK|^\text{SM}$ obtained using lattice QCD inputs including
exclusive $\Vcb$ have $5.1\sigma \sim 3.9\sigma$ tension
with the experimental results for $|\epsK|^\text{Exp}$, while there is
no tension for those obtained using inclusive $\Vcb$.
%
\input{ tab_epsK_utu_bgi_indir.tex}
In Table \ref{tab:epsK:utu:bgi:direct}, we report results for
$|\epsK|^\text{SM}$ evaluated using the \texttt{direct} method for
$\xi_0$, while the rest of inputs parameters are the same as in Table
\ref{tab:epsK:utu:bgi:indirect}.
Here we find that the theoretical results for $|\epsK|^\text{SM}$
determined using lattice QCD inputs with exclusive $\Vcb$ have $4.9
\sigma \sim 3.8 \sigma$ tension with the experimental
results for $|\epsK|^\text{Exp}$.
This is consistent with what we find from Table
\ref{tab:epsK:utu:bgi:indirect}.
%
\input{ tab_epsK_utu_bgi_dir.tex}
Fig.~\ref{fig:epsK:cmp:ctu:utu:bgi} is the same kind of plot as
Fig.~\ref{fig:epsK:cmp:ctu:utu:rbc}, obtained using the \texttt{BGI}
estimate for $\xi_\text{LD}$ instead of the \texttt{RBC-UKQCD}
estimate for $\xi_\text{LD}$, while the rest of input parameters are
the same.
%
\input{ fig_epsK_ex_ctu_utu_bgi.tex}

In Table \ref{tab:err-budget:utu:bgi}, we present the error budget for
$|\epsK|^\text{SM}$ in Fig.~\ref{fig:epsK:cmp:ctu:utu:bgi}
\subref{fig:epsK-ex:ctu:bgi:traditional}.
Here we find that the dominant error ($\cong 61 \%$) in
$|\epsK|^\text{SM}$ comes from exclusive $\Vcb$, and the sub-dominant
errors come from $\bar{\eta}$ and $\eta_{tt}^\text{BGS}$.
Hence, if we are to observe $\Delta \epsK$ become much larger than $5
\sigma$, it is essential to reduce the error in exclusive $\Vcb$.
%
\input{ tab_err_bud_utu_bgi.tex}
%

%
%\wlee{\\ EDIT by wlee \\}
%

%% file: tab_epsK_ctu_rbc_indir.tex
%
% Table of input parameters
%
\begin{table}[htbp]
  \caption{ $|\epsK|$ in units of $1.0 \times 10^{-3}$ obtained using
    the FLAG-24 results for $\BK$, AOF for the Wolfenstein parameters,
    the \texttt{indirect} method for $\xi_0$, the \texttt{RBC-UKQCD}
    estimate for $\xi_\mathrm{LD}$, and the \texttt{traditional}
    method for $\eta_i$ of $c-t$ unitarity. The abbreviation excl
    (incl) represents exclusive (inclusive) $\Vcb$. The abbreviation
    exp represents the experimental result for $|\epsK|$:
    $|\epsK|^\text{Exp}$. The labels in the source column match those
    in Table \ref{tab:Vcb}.}
  \label{tab:epsK:ctu:rbc:indirect}
  \renewcommand{\arraystretch}{1.4}
  \begin{ruledtabular}
    \begin{tabular}{l l l l r}
      $\Vcb$    & method   & source       & $|\epsK|^\text{SM}$ & $\Delta\epsK/\sigma$
      \\ \hline
      %excl      & BGL      & FNAL/MILC-22 & $1.461 \pm 0.152$   & $5.02$
      excl      & BGL      & FNAL/MILC-22 & $1.453 \pm 0.152$   & $5.10$
      \\
      %excl      & comb     & HFLAV-23     & $1.560 \pm 0.133$   & $5.01$
      excl      & comb     & HFLAV-23     & $1.551 \pm 0.132$   & $5.10$
      \\
      %excl      & comb     & FLAG-24      & $1.625 \pm 0.140$   & $4.29$
      excl      & comb     & FLAG-24      & $1.605 \pm 0.138$   & $4.50$
      \\
      %excl      & comb     & HPQCD-23     & $1.553 \pm 0.169$   & $3.98$
      excl      & comb     & HPQCD-23     & $1.544 \pm 0.168$   & $4.06$
      \\ \hline
      %incl      & 1S       & 1S-23        & $2.029 \pm 0.155$   & $1.28$
      incl      & 1S       & 1S-23        & $2.017 \pm 0.154$   & $1.36$
      \\
      %incl      & kinetic  & Gambino-21   & $2.061 \pm 0.163$   & $1.02$
      incl      & kinetic  & Gambino-21   & $2.050 \pm 0.162$   & $1.10$
      \\ \hline\hline
      $|\epsK|^\text{Exp}$
                & exp      & PDG-24       & $2.228 \pm 0.011$   & $0.00$
    \end{tabular}
  \end{ruledtabular}
\end{table}

%% file: tab_epsK_ctu_rbc_dir.tex
%
% Table of input parameters
%
\begin{table}[htbp]
  \caption{ $|\epsK|$ in units of $1.0 \times 10^{-3}$ obtained using
    the \texttt{direct} method for $\xi_0$, while the rest of the
    input parameters are the same as in Table
    \ref{tab:epsK:ctu:rbc:indirect}. The notation is the same as in
    Table \ref{tab:epsK:ctu:rbc:indirect}.}
  \label{tab:epsK:ctu:rbc:direct}
  \renewcommand{\arraystretch}{1.4}
  \begin{ruledtabular}
    \begin{tabular}{l l l l r}
      $\Vcb$ & method & source       & $|\epsK|^\text{SM}$ & $\Delta\epsK/\sigma$
      \\ \hline
      %excl   & BGL    & FNAL/MILC-22 & $1.436 \pm 0.155$   & $5.08$
      excl   & BGL    & FNAL/MILC-22 & $1.428 \pm 0.155$   & $5.17$
      \\
      %excl   & comb   & HFLAV-23     & $1.535 \pm 0.136$   & $5.07$
      excl   & comb   & HFLAV-23     & $1.526 \pm 0.136$   & $5.16$
      \\
      %excl   & comb   & FLAG-24      & $1.600 \pm 0.143$   & $4.37$
      excl   & comb   & FLAG-24      & $1.580 \pm 0.141$   & $4.58$
      \\
      %excl   & comb   & HPQCD-23     & $1.528 \pm 0.172$   & $4.07$
      excl   & comb   & HPQCD-23     & $1.519 \pm 0.171$   & $4.14$
      \\ \hline\hline
      $|\epsK|^\mathrm{Exp}$
             & exp    & PDG-24       & $2.228 \pm 0.011$   & $0.00$
    \end{tabular}
  \end{ruledtabular}
\end{table}

%% file: fig_epsK_in_ex_ctu_rbc.tex
%
% file = fig_epsK_in_ex_ctu_rbc.tex
%
\begin{figure*}[htbp]
  \renewcommand{\subfigcapskip}{0.5em}
  \renewcommand{\subfiglabelskip}{0.3em}
  \subfigure[Exclusive $\Vcb$ (FNAL/MILC-22)]{
    \includegraphics[width=0.47\linewidth]
       {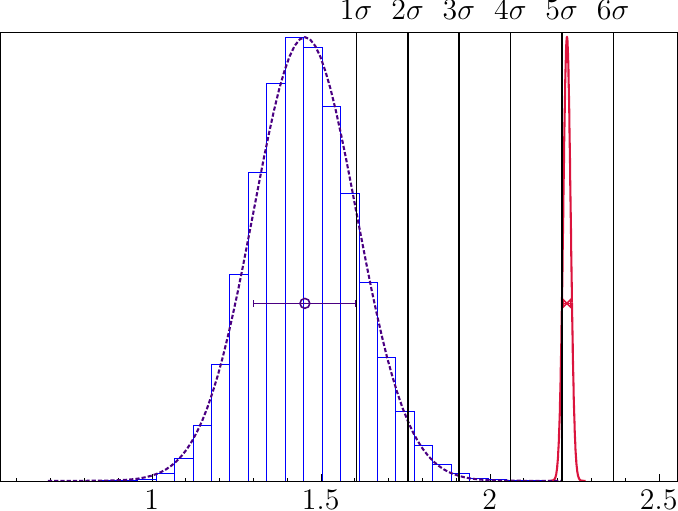}
    \label{fig:epsK-ex:ctu:rbc:indirect}
    }
  \hfill
  \subfigure[Inclusive $\Vcb$ (1S-23)]{
    \includegraphics[width=0.47\linewidth]
                    {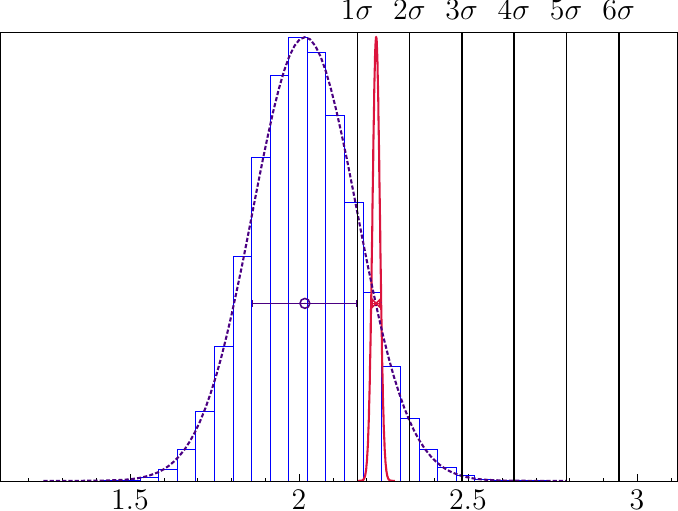}
    \label{fig:epsK-in:ctu:rbc:indirect}
    }
  \caption{$|\epsK|^\text{SM}$ with
    \subref{fig:epsK-ex:ctu:rbc:indirect} exclusive $\Vcb$ (left) and
    \subref{fig:epsK-in:ctu:rbc:indirect} inclusive $\Vcb$ (right) in
    units of $1.0\times 10^{-3}$. Here, we use the FLAG-24 results for
    $\BK$, the AOF results for the Wolfenstein parameters, the
    \texttt{indirect} method for $\xi_0$, the \texttt{RBC-UKQCD}
    estimate for $\xi_\text{LD}$, and the \texttt{traditional} method
    for $\eta_i$ of $c-t$ unitarity. Here the exclusive $\Vcb$
    represents FNAL/MILC-22 in Table \ref{tab:Vcb}
    \subref{tab:ex-Vcb}, and the inclusive $\Vcb$ corresponds to 1S-23
    in Table \ref{tab:Vcb} \subref{tab:in-Vcb}.  The red curve
    represents the experimental results for $|\epsK|^\text{Exp}$ and
    the blue curve represents the theoretical results for
    $|\epsK|^\text{SM}$ calculated directly from the SM. }
  \label{fig:epsK:cmp:ctu:rbc:indirect}
\end{figure*}

%% file: fig_epsK_history_ctu_rbc.tex
%
% file = fig_epsK_history_ctu_rbc.tex
%
\begin{figure*}[htbp]
  \renewcommand{\subfigcapskip}{0.5em}
  \renewcommand{\subfiglabelskip}{0.3em}
  \subfigure[Time evolution of $\Delta \epsK/\sigma$]{
    \includegraphics[width=0.495\linewidth]
                    {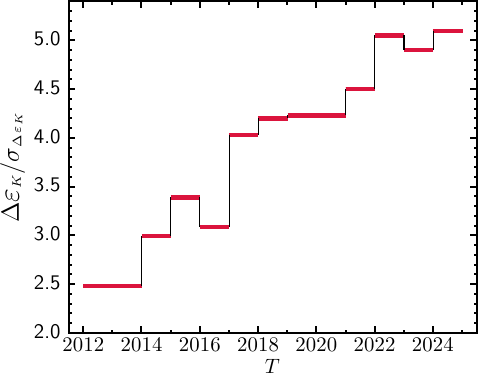}
    \label{fig:depsK:ctu:rbc:his}
    }
  \hfill
  \subfigure[Time evolution of the average and error of $\Delta\epsK$]{
    \includegraphics[width=0.455\linewidth]
                    {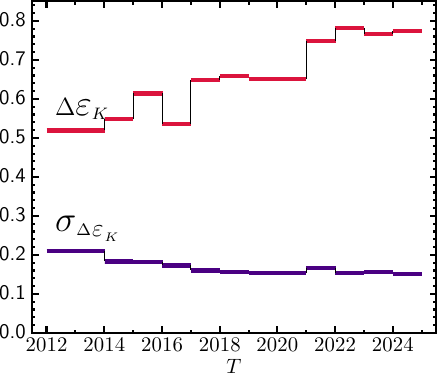}
    \label{fig:depsK+sigma:ctu:rbc:his}
    }
  \caption{ Time history of \subref{fig:depsK:ctu:rbc:his}
    $\Delta\epsK/\sigma$, and \subref{fig:depsK+sigma:ctu:rbc:his}
    $\Delta\epsK$ and $\sigma_{\Delta\epsK}$.  We define $\Delta \epsK
    \equiv |\epsK|^\text{Exp} - |\epsK|^\text{SM}$.  Here $\sigma =
    \sigma_{\Delta\epsK}$ represents the error of $\Delta\epsK$.  The
    $\Delta\epsK$ is obtained using the same input parameters as in
    Fig.~\ref{fig:epsK:cmp:ctu:rbc:indirect}
    \subref{fig:epsK-ex:ctu:rbc:indirect}. }
  \label{fig:depsK:sum:ctu:rbc:his}
\end{figure*}

%% file: tab_err_bud_ctu_rbc.tex
%
% file = tab_err_bud_ctu_rbc.tex
%
%--------------
% Error budget
%--------------
\begin{table}[tb!]
  \caption{ Error budget for $|\epsK|^\text{SM}$ obtained using the
    AOF method for the Wolfenstein parameters, the exclusive $\Vcb$
    (FNAL/MILC-22), the FLAG-24 $\BK$, the RBC-UKQCD estimate for
    $\xi_\text{LD}$, the indirect method for $\xi_0$, and the
    traditional method for $\eta_i$ of $c-t$ unitarity. Here the
    values are fractional contributions to the total error obtained
    using the formula in Ref.~\cite{ Bailey:2015tba}. }
  \label{tab:err-budget:ctu:rbc}
  \renewcommand{\arraystretch}{1.3}
  \begin{ruledtabular}
  \begin{tabular}{c l l}
    source          & error (\%)            & memo \\
    \hline
    $\Vcb$          &           51.9        & exclusive \\ 
    $\eta_{ct}$     &           21.9        & $c-t$ Box \\ \hline
    $\bar{\eta}$    & \phantom{0}9.4        & AOF \\
    $\eta_{cc}$     & \phantom{0}9.3        & $c-c$ Box \\
    $\xi_\text{LD}$ & \phantom{0}2.2        & RBC/UKQCD \\
    $\bar{\rho}$    & \phantom{0}2.0        & AOF \\
    $\hat{B}_K$     & \phantom{0}1.6        & FLAG-24 \\ \hline
    $\eta_{tt}$     & \phantom{0}0.69       & $t-t$ Box \\
    $\xi_0$         & \phantom{0}0.65       & RBC/UKQCD \\
    $F_K$           & \phantom{0}0.16       & FLAG-24 \\
    $\lambda$       & \phantom{0}0.15       & FLAG-24 \\ \hline
    $m_t$           & \phantom{0}0.055      & $m_t(m_t)$ \\
    $m_c$           & \phantom{0}0.053      & $m_c(m_c)$ \\
    $\vdots$        & \phantom{00}$\vdots$  & $\vdots$
  \end{tabular}
  \end{ruledtabular}
\end{table}

%% file: tab_epsK_ctu_bgi_indir.tex
%
% Table of input parameters
%
\begin{table}[htbp]
  \caption{ $|\epsK|$ in units of $1.0 \times 10^{-3}$ obtained using
    the \texttt{indirect} method for $\xi_0$, the \texttt{BGI}
    estimate for $\xi_\mathrm{LD}$, and the \texttt{traditional}
    method for $\eta_i$ of $c-t$ unitarity, while the rest of the
    input parameters are the same as in Table
    \ref{tab:epsK:ctu:rbc:indirect}. The notation is the same as in
    Table \ref{tab:epsK:ctu:rbc:indirect}. }
  \label{tab:epsK:ctu:bgi:indirect}
  \renewcommand{\arraystretch}{1.4}
  \begin{ruledtabular}
    \begin{tabular}{l l l l r}
      $\Vcb$ & method  & source       & $|\epsK|^\text{SM}$ & $\Delta\epsK/\sigma$   \\ \hline
      %excl   & BGL     & FNAL/MILC-22 & $1.509 \pm 0.155$   & $4.62$
      excl   & BGL     & FNAL/MILC-22 & $1.501 \pm 0.154$   & $4.70$
      \\
      %excl   & comb    & HFLAV-23     & $1.608 \pm 0.136$   & $4.56$
      excl   & comb    & HFLAV-23     & $1.599 \pm 0.135$   & $4.64$
      \\
      %excl   & comb    & FLAG-24      & $1.673 \pm 0.143$   & $3.88$
      excl   & comb    & FLAG-24      & $1.653 \pm 0.140$   & $4.08$
      \\
      %excl   & comb    & HPQCD-23     & $1.601 \pm 0.171$   & $3.65$
      excl   & comb    & HPQCD-23     & $1.592 \pm 0.171$   & $3.73$
      \\ \hline
      %incl   & 1S      & 1S-23        & $2.077 \pm 0.156$   & $0.97$
      incl   & 1S      & 1S-23        & $2.065 \pm 0.155$   & $1.05$
      \\
      %incl   & kinetic & Gambino-21   & $2.109 \pm 0.164$   & $0.72$
      incl   & kinetic & Gambino-21   & $2.098 \pm 0.163$   & $0.80$
      \\ \hline\hline
      $|\epsK|^\mathrm{Exp}$
             & exp     & PDG-24       & $2.228 \pm 0.011$   & $0.00$
    \end{tabular}
  \end{ruledtabular}
\end{table}

%% file: tab_epsK_ctu_bgi_dir.tex
%
% Table of input parameters
%
\begin{table}[htbp]
  \caption{ $|\epsK|$ in units of $1.0 \times 10^{-3}$ obtained using
    the \texttt{direct} method for $\xi_0$, the \texttt{BGI} estimate
    for $\xi_\mathrm{LD}$, and the \texttt{traditional} method for
    $\eta_i$ of $c-t$ unitarity, while the rest of the input
    parameters are the same as in Table
    \ref{tab:epsK:ctu:rbc:indirect}. The notation is the same as in
    Table \ref{tab:epsK:ctu:rbc:indirect}.}
  \label{tab:epsK:ctu:bgi:direct}
  \renewcommand{\arraystretch}{1.4}
  \begin{ruledtabular}
    \begin{tabular}{l l l l r}
      $\Vcb$ & method & source       & $|\epsK|^\text{SM}$ & $\Delta\epsK/\sigma$
      \\ \hline
      excl   & BGL    & FNAL/MILC-22 & $1.486 \pm 0.160$   & $4.64$
      \\
      excl   & comb   & HFLAV-23     & $1.584 \pm 0.141$   & $4.56$
      \\
      excl   & comb   & FLAG-23      & $1.638 \pm 0.146$   & $4.03$
      \\
      excl   & comb   & HPQCD-23     & $1.576 \pm 0.175$   & $3.71$
      \\ \hline\hline
      $|\epsK|^\mathrm{Exp}$
             & exp    & PDG-24       & $2.228 \pm 0.011$   & $0.00$
    \end{tabular}
  \end{ruledtabular}
\end{table}

%% file: fig_epsK_in_ex_ctu_bgi.tex
%
% file = fig_epsK_in_ex_ctu_bgi.tex
%
\begin{figure*}[htbp]
\renewcommand{\subfigcapskip}{0.5em}
\renewcommand{\subfiglabelskip}{0.3em}
  \subfigure[Exclusive $\Vcb$ (FNAL/MILC-22)]{
    \includegraphics[width=0.47\linewidth]
       {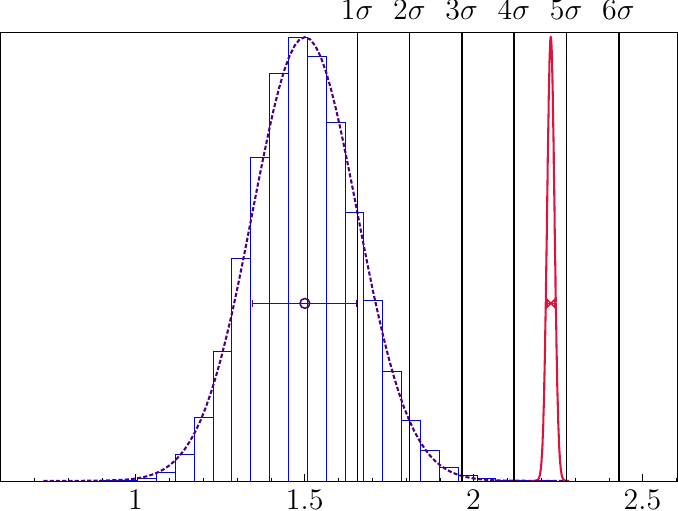}
    \label{fig:epsK-ex:ctu:bgi}
    }
  \hfill
  \subfigure[Inclusive $\Vcb$ (1S-23)]{
    \includegraphics[width=0.47\linewidth]
                    {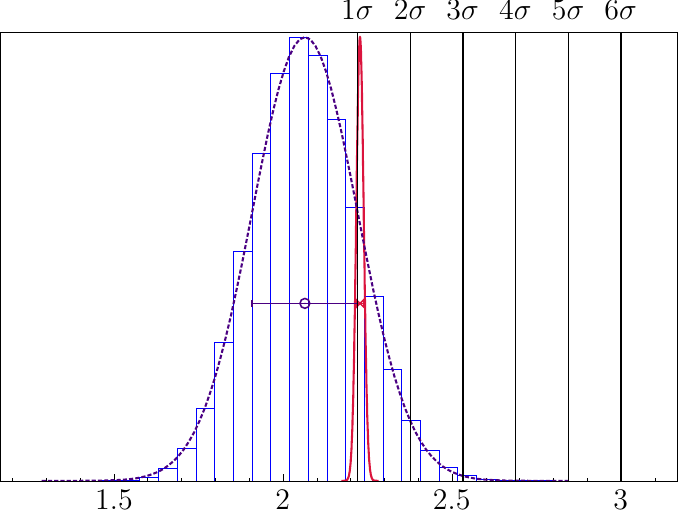}
    \label{fig:epsK-in:ctu:bgi}
    }
  \caption{$|\epsK|$ with \subref{fig:epsK-ex:ctu:bgi} exclusive
    $\Vcb$ (left) and \subref{fig:epsK-in:ctu:bgi} inclusive $\Vcb$
    (right) in units of $1.0\times 10^{-3}$. Here we use the BGI
    estimate in Eq.~\eqref{eq:xiLD:BGI} for $\xi_\text{LD}$ with the
    rest of the input parameters set to the same as in
    Fig.~\ref{fig:epsK:cmp:ctu:rbc:indirect}. The red curve represents
    the experimental results for $|\epsK|^\text{Exp}$, and the blue
    curve represents the theoretical results for $|\epsK|^\text{SM}$.
  }
  \label{fig:epsK:cmp:ctu:bgi}
\end{figure*}

%% file: fig_epsK_history_ctu_bgi.tex
%
% file = fig_epsK_history_ctu_bgi.tex
%
\begin{figure*}[htbp]
  \renewcommand{\subfigcapskip}{0.5em}
  \renewcommand{\subfiglabelskip}{0.3em}
  \subfigure[Time evolution of $\Delta \epsK/\sigma$]{
    \includegraphics[width=0.491\linewidth]
                    {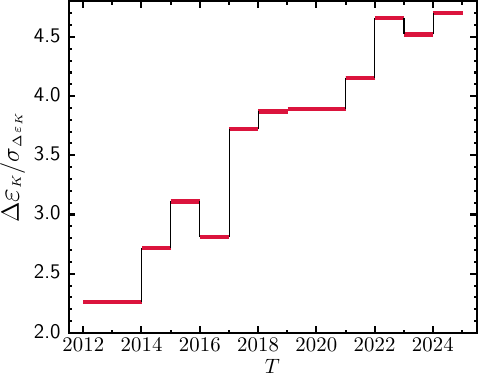}
    \label{fig:depsK:ctu:bgi:his}
    }
  \hfill
  \subfigure[Time evolution of  the average and error of $\Delta\epsK$]{
    \includegraphics[width=0.45\linewidth]
                    {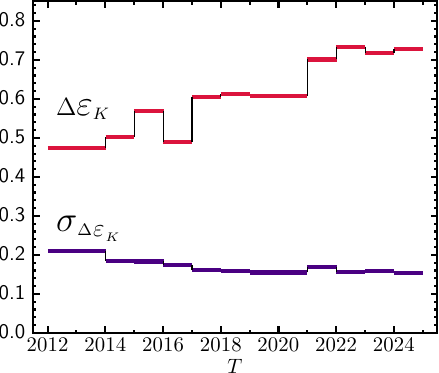}
    \label{fig:depsK+sigma:ctu:bgi:his}
    }
  \caption{ Time history of \subref{fig:depsK:ctu:bgi:his}
    $\Delta\epsK/\sigma$, and \subref{fig:depsK+sigma:ctu:bgi:his}
    $\Delta\epsK$ and $\sigma_{\Delta\epsK}$. Here $\sigma =
    \sigma_{\Delta\epsK}$. The $\Delta\epsK$ is obtained using the
    same input parameters as in Fig.~\ref{fig:epsK:cmp:ctu:bgi}
    \subref{fig:epsK-ex:ctu:bgi}. }
  \label{fig:depsK:sum:ctu:bgi:his}
\end{figure*}

%% file: tab_err_bud_ctu_bgi.tex
%
% file = tab_err_bud_ctu_bgi.tex
%
%--------------
% Error budget
%--------------
\begin{table}[tb!]
  \caption{ Error budget for $|\epsK|^\text{SM}$ obtained using the
    AOF results for the Wolfenstein parameters, the FNAL/MILC-22
    results for exclusive $\Vcb$, the FLAG-2017 results for $\BK$, the
    \texttt{indirect} method for $\xi_0$, the \texttt{BGI} estimate
    for $\xi_\text{LD}$, and the \texttt{traditional} method for
    $\eta_i$ of $c-t$ unitarity. Here, the values are fractional
    contributions to the total error obtained using the formula in
    Ref.~\cite{ Bailey:2015tba}. }
  \label{tab:err-budget:ctu:bgi}
  \renewcommand{\arraystretch}{1.3}
  \begin{ruledtabular}
  \begin{tabular}{c l l}
    source          & error (\%)  & memo \\
    \hline
    $\Vcb$          &           50.1        & exclusive \\
    $\eta_{ct}$     &           21.2        & $c-t$ Box \\ \hline
    $\bar{\eta}$    & \phantom{0}9.0        & AOF \\
    $\eta_{cc}$     & \phantom{0}9.0        & $c-c$ Box \\
    $\xi_\text{LD}$ & \phantom{0}5.5        & BGI estimate \\
    $\bar{\rho}$    & \phantom{0}1.9        & AOF \\
    $\hat{B}_K$     & \phantom{0}1.5        & FLAG \\ \hline
    $\eta_{tt}$     & \phantom{0}0.67       & $t-t$ Box \\
    $\xi_0$         & \phantom{0}0.63       & RBC/UKQCD \\
    $F_K$           & \phantom{0}0.15       & $m_t(m_t)$ \\
    $\lambda$       & \phantom{0}0.14       & $\Vus$ \\ \hline
    $m_t$           & \phantom{0}0.053      & $m_t(m_t)$ \\
    $m_c$           & \phantom{0}0.051      & $m_c(m_c)$ \\
    $\vdots$        & \phantom{00}$\vdots$  & $\vdots$
  \end{tabular}
  \end{ruledtabular}
\end{table}

%% file: tab_epsK_utu_rbc_indir.tex
%
% Table of input parameters
%
\begin{table*}[htbp]
  \caption{ $|\epsK|$ in units of $1.0 \times 10^{-3}$ obtained using
    the FLAG-24 results for $\BK$, the AOF results for the Wolfenstein
    parameters, the \texttt{indirect} method for $\xi_0$, the
    \texttt{RBC-UKQCD} estimate for $\xi_\mathrm{LD}$, and the
    \texttt{BGS} method for $\eta_i$ ($=\eta_i^\text{BGS}$) of $u-t$
    unitarity. Here $\sigma_1^\text{BGS}$, $\delta\epsK^\text{BGS}$,
    and $\sigma_t^\text{BGS}$ are explained in
    Eq.~\eqref{eq:BGS:sigma_t-1}. The rest of notations are the same
    as Table \ref{tab:epsK:ctu:rbc:indirect}. The labels in the source
    column match those in Table \ref{tab:Vcb}.  }
  \label{tab:epsK:utu:rbc:indirect}
  \renewcommand{\arraystretch}{1.4}
  \begin{ruledtabular}
    \begin{tabular}{l l l l l l l r}
      $\Vcb$    & method   & source       & $|\epsK|^\text{SM}$
      & $\sigma_1^\text{BGS}$ & $\delta\epsK^\text{BGS}$
      & $\sigma_t^\text{BGS}$ & $\Delta\epsK/\sigma$
      \\ \hline
      %excl      & BGL      & FNAL/MILC-22 & 1.493 & 0.133 & 0.032 & 0.137
      excl      & BGL      & FNAL/MILC-22 & 1.484 & 0.133 & 0.032 & 0.137
      & 5.43
      \\
      %excl      & comb     & HFLAV-23     & 1.592 & 0.110 & 0.031 & 0.114
      excl      & comb     & HFLAV-23     & 1.582 & 0.110 & 0.031 & 0.114
      & 5.65
      \\
      %excl      & comb     & FLAG-24      & 1.656 & 0.118 & 0.031 & 0.122
      excl      & comb     & FLAG-24      & 1.635 & 0.116 & 0.030 & 0.120
      & 4.93
      \\
      %excl      & comb     & HPQCD-23     & 1.584 & 0.152 & 0.031 & 0.155
      excl      & comb     & HPQCD-23     & 1.575 & 0.151 & 0.031 & 0.154
      & 4.24
      \\ \hline
      %incl      & 1S       & 1S-23        & 2.055 & 0.132 & 0.026 & 0.135
      incl      & 1S       & 1S-23        & 2.043 & 0.132 & 0.026 & 0.134
      & 1.37
      \\
      %incl      & kinetic  & Gambino-21   & 2.087 & 0.141 & 0.026 & 0.143
      incl      & kinetic  & Gambino-21   & 2.075 & 0.140 & 0.025 & 0.142
      & 1.07
      \\ \hline
      $|\epsK|^\text{Exp}$
      & exp      & PDG-24                 & 2.228 &       &       & 0.011
      & 0.00
    \end{tabular}
  \end{ruledtabular}
\end{table*}

%% file: tab_epsK_utu_rbc_dir.tex
%
% Table of input parameters
%
\begin{table}[htbp]
  \caption{ $|\epsK|$ in units of $1.0 \times 10^{-3}$ obtained using
    the \texttt{direct} method for $\xi_0$, and the \texttt{BGS}
    method for $\eta_i$ of $u-t$ unitarity, while the rest of the
    input parameters are the same as in Table
    \ref{tab:epsK:utu:rbc:indirect}. Here the error of
    $|\epsK|^\text{SM}$ represents the total error
    $\sigma_t^\text{BGS}$ in Eq.~\eqref{eq:BGS:sigma_t-1}. The
    notation is the same as in Table \ref{tab:epsK:utu:rbc:indirect}.
    }
  \label{tab:epsK:utu:rbc:direct}
  \renewcommand{\arraystretch}{1.4}
  \begin{ruledtabular}
    \begin{tabular}{l l l l r}
      $\Vcb$ & method & source       & $|\epsK|^\text{SM}$ & $\Delta\epsK/\sigma$
      \\ \hline
      excl   & BGL    & FNAL/MILC-22 & $1.459 \pm 0.140$   & 5.48
      \\
      excl   & comb   & HFLAV-23     & $1.557 \pm 0.118$   & 5.68
      \\
      excl   & comb   & FLAG-24      & $1.610 \pm 0.123$   & 4.99
      \\
      excl   & comb   & HPQCD-23     & $1.550 \pm 0.157$   & 4.32
      \\ \hline
      $|\epsK|^\mathrm{Exp}$
             & exp    & PDG-24       & $2.228 \pm 0.011$   & 0.00
    \end{tabular}
  \end{ruledtabular}
\end{table}

%% file: fig_epsK_ex_ctu_utu_rbc.tex
%
% file = fig_epsK_in_ex_ctu_rbc.tex
%
\begin{figure*}[htbp]
  \renewcommand{\subfigcapskip}{0.5em}
  \renewcommand{\subfiglabelskip}{0.3em}
  \subfigure[The tradtional method ($\eta_i$ with $c-t$ unitarity)]{
    \includegraphics[width=0.47\linewidth]
    {epsK-FLAG-ex-BtoDstar-FNAL-BGL-AOF-RBC-UKQCD-indir-20241124-v1.pdf}
    \label{fig:epsK-ex:ctu:rbc:traditional}
    }
  \hfill
  \subfigure[The BGS method ($\eta_i$ with $u-t$ unitarity)]{
    \includegraphics[width=0.47\linewidth]
    {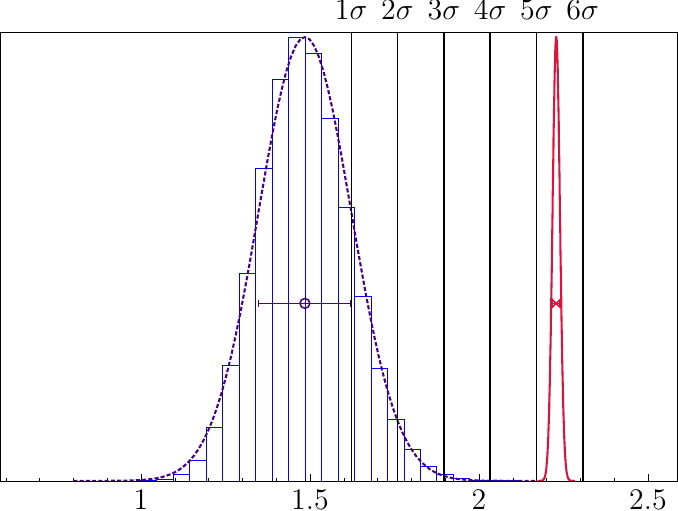}
    \label{fig:epsK-ex:utu:rbc:BGS}
    }
  \caption{$|\epsK|^\text{SM}$ with the exclusive $\Vcb$ in units of
    $1.0\times 10^{-3}$, using
    \subref{fig:epsK-ex:ctu:rbc:traditional} $\eta_i$ with $c-t$
    unitarity and \subref{fig:epsK-ex:utu:rbc:BGS} $\eta_i$ with $u-t$
    unitarity.  Here we use the FLAG-24 results for $\BK$, the AOF
    results for the Wolfenstein parameters, the \texttt{indirect}
    method for $\xi_0$, the \texttt{RBC-UKQCD} estimate for
    $\xi_\text{LD}$. Here the exclusive $\Vcb$ indicates FNAL/MILC-22
    in Table \ref{tab:Vcb} \subref{tab:ex-Vcb}.  The red curve
    represents $|\epsK|^\text{Exp}$, and
    the blue curve represents $|\epsK|^\text{SM}$.  }
  \label{fig:epsK:cmp:ctu:utu:rbc}
\end{figure*}

%% file: tab_err_bud_utu_rbc.tex
%
% file = tab_err_bud_utu_rbc.tex
%
%--------------
% Error budget
%--------------
\begin{table}[tb!]
  \caption{ Error budget for $|\epsK|^\text{SM}$ obtained using the
    AOF method for the Wolfenstein parameters, the exclusive $\Vcb$
    (FNAL/MILC-22), the FLAG-24 $\BK$, the RBC-UKQCD estimate for
    $\xi_\text{LD}$, the indirect method for $\xi_0$, and the BGS
    method for $\eta_i$ of $u-t$ unitarity. Here the values are
    fractional contributions to the total error obtained using the
    formula in Ref.~\cite{ Bailey:2015tba}.
  }
  \label{tab:err-budget:utu:rbc}
  \renewcommand{\arraystretch}{1.3}
  \begin{ruledtabular}
  \begin{tabular}{c l l}
    source          & error (\%)            & memo \\
    \hline
    $\Vcb$                 &           63.1        & exclusive \\ 
    $\bar{\eta}$           &           12.0        & AOF \\
    $\eta_{tt}^\BGS$       &           10.7        & BGS \\ \hline
    $\delta\epsK^\BGS$     & \phantom{0}5.4        & CVM \\
    $\xi_\text{LD}$        & \phantom{0}2.9        & RBC/UKQCD \\
    $\bar{\rho}$           & \phantom{0}2.2        & AOF \\
    $\hat{B}_K$            & \phantom{0}2.0        & FLAG-24 \\ \hline
    $\xi_0$                & \phantom{0}0.80       & RBC/UKQCD \\
    $\eta_{ut}^\BGS$       & \phantom{0}0.28       & BGS  \\
    $F_K$                  & \phantom{0}0.21       & FLAG-24 \\
    $\lambda$              & \phantom{0}0.19       & FLAG-24 \\
    $m_c$                  & \phantom{0}0.10       & $m_c(m_c)$ \\ \hline
    $m_t$                  & \phantom{0}0.061      & $m_t(m_t)$ \\
    $\vdots$               & \phantom{00}$\vdots$  & $\vdots$
  \end{tabular}
  \end{ruledtabular}
\end{table}

%% file: tab_epsK_utu_bgi_indir.tex
%
% Table of input parameters
%
\begin{table}[htbp]
  \caption{ $|\epsK|$ in units of $1.0 \times 10^{-3}$ obtained using
    the \texttt{indirect} method for $\xi_0$, the \texttt{BGI}
    estimate for $\xi_\mathrm{LD}$, and the \texttt{BGS} method for
    $\eta_i$ of $u-t$ unitarity, while the rest of the input parameters
    are the same as in Table \ref{tab:epsK:utu:rbc:indirect}. The
    notation is the same as in Table
    \ref{tab:epsK:utu:rbc:indirect}.
  }
  \label{tab:epsK:utu:bgi:indirect}
  \renewcommand{\arraystretch}{1.4}
  \begin{ruledtabular}
    \begin{tabular}{l l l l r}
      $\Vcb$ & method  & source       & $|\epsK|^\text{SM}$ & $\Delta\epsK/\sigma$   \\ \hline
      %excl   & BGL     & FNAL/MILC-22 & $1.541 \pm 0.136$   & $5.02$
      excl   & BGL     & FNAL/MILC-22 & $1.532 \pm 0.139$   & $4.97$
      \\
      %excl   & comb    & HFLAV-23     & $1.639 \pm 0.113$   & $5.18$
      excl   & comb    & HFLAV-23     & $1.630 \pm 0.117$   & $5.10$
      \\
      %excl   & comb    & FLAG-24      & $1.703 \pm 0.121$   & $4.33$
      excl   & comb    & FLAG-24      & $1.683 \pm 0.122$   & $4.43$
      \\
      %excl   & comb    & HPQCD-23     & $1.632 \pm 0.154$   & $3.87$
      excl   & comb    & HPQCD-23     & $1.622 \pm 0.156$   & $3.87$
      \\ \hline
      %incl   & 1S      & 1S-23        & $2.103 \pm 0.133$   & $0.94$
      incl   & 1S      & 1S-23        & $2.091 \pm 0.135$   & $1.01$
      \\
      %incl   & kinetic & Gambino-21   & $2.135 \pm 0.142$   & $0.65$
      incl   & kinetic & Gambino-21   & $2.123 \pm 0.143$   & $0.73$
      \\ \hline\hline
      $|\epsK|^\mathrm{Exp}$
             & exp     & PDG-24       & $2.228 \pm 0.011$   & $0.00$
    \end{tabular}
  \end{ruledtabular}
\end{table}

%% file: tab_epsK_utu_bgi_dir.tex
%
% Table of input parameters
%
\begin{table}[htbp]
  \caption{ $|\epsK|$ in units of $1.0 \times 10^{-3}$ obtained using
    the \texttt{direct} method for $\xi_0$, the \texttt{BGI} estimate
    for $\xi_\mathrm{LD}$, and the \texttt{BGS} method for $\eta_i$ of
    $u-t$ unitarity, while the rest of the input parameters are the
    same as in Table \ref{tab:epsK:utu:rbc:indirect}. The notation is
    the same as in Table \ref{tab:epsK:utu:rbc:indirect}.
  }
  \label{tab:epsK:utu:bgi:direct}
  \renewcommand{\arraystretch}{1.4}
  \begin{ruledtabular}
    \begin{tabular}{l l l l r}
      $\Vcb$ & method & source       & $|\epsK|^\text{SM}$ & $\Delta\epsK/\sigma$
      \\ \hline
      %excl   & BGL    & FNAL/MILC-22 & $1.526 \pm 0.142$   & $4.92$
      excl   & BGL    & FNAL/MILC-22 & $1.517 \pm 0.145$   & $4.88$
      \\
      %excl   & comb   & HFLAV-23     & $1.624 \pm 0.120$   & $5.00$
      excl   & comb   & HFLAV-23     & $1.615 \pm 0.124$   & $4.94$
      \\
      %excl   & comb   & FLAG-24      & $1.688 \pm 0.127$   & $4.22$
      excl   & comb   & FLAG-24      & $1.668 \pm 0.129$   & $4.33$
      \\
      %excl   & comb   & HPQCD-23     & $1.617 \pm 0.159$   & $3.84$
      excl   & comb   & HPQCD-23     & $1.607 \pm 0.161$   & $3.84$
      \\ \hline\hline
      $|\epsK|^\mathrm{Exp}$
             & exp    & PDG-24       & $2.228 \pm 0.011$   & $0.00$
    \end{tabular}
  \end{ruledtabular}
\end{table}

%% file: fig_epsK_ex_ctu_utu_bgi.tex
%
% file = fig_epsK_in_ex_utu_bgi.tex
%
\begin{figure*}[htbp]
  \renewcommand{\subfigcapskip}{0.5em}
  \renewcommand{\subfiglabelskip}{0.3em}
  \subfigure[The tradtional method ($\eta_i$ with $c-t$ unitarity)]{
    \includegraphics[width=0.47\linewidth]
    {epsK-FLAG-ex-BtoDstar-FNAL-BGL-AOF-BGI-indir-20241124-v1.pdf}
    \label{fig:epsK-ex:ctu:bgi:traditional}
    }
  \hfill
  \subfigure[The BGS method ($\eta_i$ with $u-t$ unitarity)]{
    \includegraphics[width=0.47\linewidth]
    {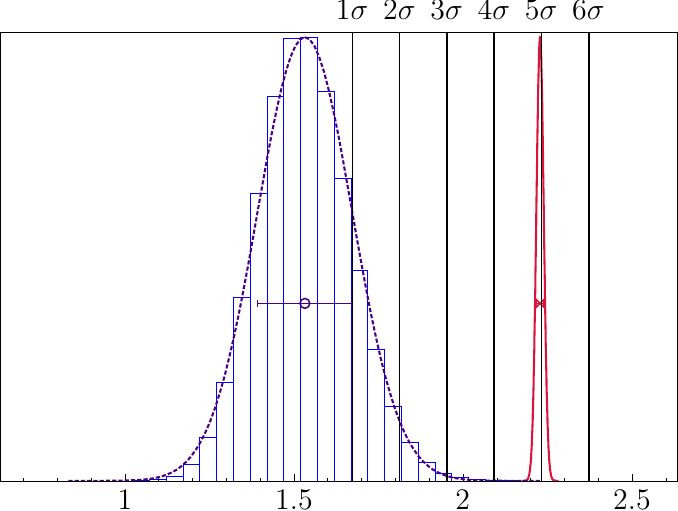}
    \label{fig:epsK-ex:utu:bgi:BGS}
    }
  \caption{$|\epsK|$ with exclusive $\Vcb$ in units of $1.0\times
    10^{-3}$, using \subref{fig:epsK-ex:ctu:bgi:traditional} $\eta_i$
    with $c-t$ unitarity, and \subref{fig:epsK-ex:utu:bgi:BGS}
    $\eta_i$ with $u-t$ unitarity.  Here we use the \texttt{BGI}
    estimate for $\xi_\text{LD}$, while the rest of the input
    parameters are the same as in
    Fig.~\ref{fig:epsK:cmp:ctu:utu:rbc}. The red (blue) curve
    represents $|\epsK|^\text{Exp}$ ($|\epsK|^\text{SM}$).  }
  \label{fig:epsK:cmp:ctu:utu:bgi}
\end{figure*}

%% file: tab_err_bud_utu_bgi.tex
%
% file = tab_err_bud_utu_bgi.tex
%
%--------------
% Error budget
%--------------
\begin{table}[tb!]
  \caption{ Error budget for $|\epsK|^\text{SM}$ obtained using the
    AOF method for the Wolfenstein parameters, the exclusive $\Vcb$,
    the FLAG-2017 $\BK$, the indirect method for $\xi_0$, the BGI
    estimate for $\xi_\text{LD}$, and the BGS method for $\eta_i$ of
    $u-t$ unitarity. Here, the values are fractional contributions to
    the total error obtained using the formula in Ref.~\cite{
      Bailey:2015tba}.
  }
  \label{tab:err-budget:utu:bgi}
  \renewcommand{\arraystretch}{1.3}
  \begin{ruledtabular}
  \begin{tabular}{c l l}
    source                 & error (\%)  & memo \\
    \hline
    $\Vcb$                 &           60.5        & exclusive \\
    $\bar{\eta}$           &           11.5        & AOF \\
    $\eta_{tt}^\BGS$       &           10.3        & BGS \\ \hline
    $\xi_\text{LD}$        & \phantom{0}6.8        & RBC/UKQCD \\
    $\delta\epsK^\BGS$     & \phantom{0}5.2        & CVM \\
    $\bar{\rho}$           & \phantom{0}2.1        & AOF \\
    $\hat{B}_K$            & \phantom{0}1.9        & FLAG-24 \\ \hline
    $\xi_0$                & \phantom{0}0.77       & RBC/UKQCD \\
    $\eta_{ut}^\BGS$       & \phantom{0}0.27       & BGS  \\
    $F_K$                  & \phantom{0}0.20       & FLAG-24 \\
    $\lambda$              & \phantom{0}0.18       & FLAG-24 \\
    $m_c$                  & \phantom{0}0.10       & $m_c(m_c)$ \\ \hline
    $m_t$                  & \phantom{0}0.058      & $m_t(m_t)$ \\
    $\vdots$               & \phantom{00}$\vdots$  & $\vdots$
  \end{tabular}
  \end{ruledtabular}
\end{table}
%
%
%

%% file: conclude.tex
%
% conclude.tex
%
\section{Conclusion}
\label{sec:conclude}
In this paper, we report that there exists a remarkable tension of
$5.2\sigma \sim 4.6\sigma$ in $|\epsK|$ between the experiment and the
SM theory with lattice QCD inputs.\footnote{Here we use the
FNAL/MILC-22 results for exclusive $\Vcb$, and the traditional method
for $\eta_i$ of $c-t$ unitarity.
The upper bound of the $5.2 \sigma$ tension is obtained with the
RBC-UKQCD estimate for $\xi_\text{LD}$, and the direct method for
$\xi_0$.
The lower bound of the $4.6 \sigma$ tension is obtained with the BGI
estimate for $\xi_\text{LD}$, and the direct method for $\xi_0$.
To obtain the results, we use the AOF results for the Wolfenstein
parameters, the FLAG-24 results for $\BK$, and the direct method for
$\xi_0$. \label{fn:ctu-lat-1} } %%% end of footnote
We reported $3.4\sigma$ tension in $\Delta \epsK$ in 2015 \cite{
  Bailey:2015tba} when it exceeded the $3\sigma$ level.
We updated $4.2\sigma$ tension in $\Delta \epsK$ in 2018 \cite{
  Bailey:2018feb}, when it exceeded the $4\sigma$ level.
Here we report that the tension in $\Delta\epsK$ is $5.1\sigma$ at
present, when it has exceeded the $5\sigma$ level.
We obtain this result using the same kind of input parameters as in
Refs.~\cite{ Bailey:2015tba, Bailey:2018feb}.\footnote{ Here we use
the RBC-UKQCD method for $\xi_\text{LD}$, the indirect method for
$\xi_0$, the AOF results for the Wolfenstein parameters, the
traditional method for $\eta_i$ of $c-t$ unitarity, the FNAL/MILC-22
results for exclusive $\Vcb$, and so on. }
%

%--------------
% BGS results
%--------------
Recently, the BGS method for $\eta_i$ of $u-t$ unitarity has become
available \cite{Brod:2019rzc}.
%
\begin{comment}
It is reported that $\eta_{ut}^\text{BGS}$ has significantly smaller
uncertainty, while $\eta_{tt}^\text{BGS}$ remains almost the same
as $\eta_{tt}$ in $c-t$ unitarity.
\end{comment}
%
If we adopt the BGS method, the tension in $\Delta\epsK$ increases
further to the $5.5 \sigma \sim 4.9 \sigma$ level, when we determine
$|\epsK|^\text{SM}$ using lattice QCD inputs including exclusive
$\Vcb$ (FNAL/MILC-22).
However we find that there exists a central value mismatch (CVM) in
$|\epsK|^\text{SM}$ between the $c-t$ unitarity and the $u-t$
unitarity.
The CVM comes from small and tiny approximations that they introduced
to simplify the BGS master formula for $\eta_{ut}^\text{BGS}$ and
$\eta_{tt}^\text{BGS}$.
Here we simply take into account the CVM as another error in the BGS
method.\footnote{ More elaborate and rigorous treatment (\emph{i.e.}
not a cure-all but somewhat better solution) for the CVM error needs
further investigation in the future \cite{ wlee:2024:epsK}.  }

In Table \ref{tab:DepsK}, we present how the values of $\Delta \epsK$
have changed from our previous papers in 2015 and 2018 to 2025.
Here we find that the positive shift of $\Delta \epsK$ is similar for
the inclusive and exclusive values of $\Vcb$.
This reflects the shift in other input parameters since 2015.
We also find that there is no significant tension observed yet for
$|\epsK|^\text{SM}$ with inclusive $\Vcb$, which is obtained using the
heavy quark expansion based on the quark-hadron duality.

\input{tab_depsK}
%

%----------------
% CLN versus BGL
%----------------
There was an interesting claim \cite{ Bigi:2017njr, Grinstein:2017nlq}
which had potential to resolve the issue of the inconsistency between
the exclusive and inclusive $\Vcb$.
It turns out that there is no difference in exclusive $\Vcb$ between
CLN and BGL at present.
Hence, this issue on CLN and BGL has been completely resolved by time
evolution.
%

%----------------------------------------
% physical interpretation of the results
%----------------------------------------
Here we report a strong tension of $5.2\sigma \sim 4.6\sigma$ in
$|\epsK|$ between $|\epsK|^\text{Exp}$ and $|\epsK|^\text{SM}$
determined using exclusive $\Vcb$.${}^{\ref{fn:ctu-lat-1}}$
We also report that the tension becomes even stronger at the level of
$5.5\sigma \sim 4.9\sigma$ with exclusive $\Vcb$, if we use the
\texttt{BGS} method of the $u-t$ unitarity instead of the $c-t$
unitarity.
We also report that the tension disappears with inclusive $\Vcb$.
Hence, it is crucial to provide a balanced picture of the current
status of $|\epsK|^\text{SM}$ with respect to $\Vcb$ from the
standpoint of lattice QCD.

First, we point out that the experiments for exclusive $\Vcb$ are
completely different from those for inclusive $\Vcb$.
The exclusive $\Vcb$ is obtained using the lattice QCD results for
the semileptonic form factors combined with the experimental results.
There are multiple decay channels to determine exclusive $\Vcb$ in
lattice QCD such as $\BtoD$, $\BtoDst$, $\BstoDs$, $\BstoDsst$,
$\LbtoLc$ and so on.
All the results of exclusive $\Vcb$ in multiple channels determined by
multiple collaborations in lattice QCD are consistent with one another
within statistical uncertainty.
Therefore, the exclusive $\Vcb$ results from lattice QCD are highly
reliable from the standpoint of lattice QCD as well as theoretical and
experimental particle physics.
This is a key observation for the current status of exclusive $\Vcb$. 

On the other hand, inclusive $\Vcb$ is obtained using the heavy quark
expansion based on the quark-hadron duality.
There are a number of attempts in lattice QCD to determine inclusive
$\Vcb$ such as that in Ref.~\cite{Barone:2022gkn, Barone:2023tbl}.
However, they at present belong to a category of exploratory study
rather than that of high precision measurement.
Some day in near future it might be possible to determine inclusive
$\Vcb$ using lattice QCD tools with such a high precision as exclusive
$\Vcb$.

If we take into account the above key points, a more balanced view
from lattice QCD might be the following.
\begin{enumerate}
\item The results for exclusive $\Vcb$ from lattice QCD are highly
  reliable.
\item Here it is reported that the difference in $|\epsK|$ between
  $|\epsK|^\text{Exp}$ (experiment) and $|\epsK|^\text{SM}$ (theory),
  $\Delta \epsK = |\epsK|^\text{Exp} - |\epsK|^\text{SM} $ has a
  strong tension of $5.2\sigma \sim 4.6\sigma$ level, when
  $|\epsK|^\text{SM}$ is determined using exclusive $\Vcb$
  (FNAL/MILC-22) with $\eta_i$ of $c-t$ unitarity.
  The use of $\eta_i$ of $u-t$ unitarity leads to a even stronger
  tension in $\Delta \epsK$.
\item The determination of inclusive $\Vcb$ from lattice QCD does not
  have enough precision at present.
\item If we use inclusive $\Vcb$ from heavy quark expansion based on
  the quark-hadron duality, the tension in $\Delta \epsK$ disappears.
\end{enumerate}
This indicates that the self-consistent and reliable calculation of
$|\epsK|^\text{SM}$ from lattice QCD has $\cong 5 \sigma$ tension with
$|\epsK|^\text{Exp}$.
The first principle calculation of inclusive $\Vcb$ using lattice QCD
tools does not have enough precision yet.
This weakness drives us to use inclusive $\Vcb$ obtained using heavy
quark expansion based on the quark-hadron duality as an alternative
resource, which does not come from the first principles such as
lattice QCD.

%
%\wlee{\\ EDIT by wlee \\}
%

%% file: tab_depsK.tex
\begin{table}[h]
%%  \footnotesize
  \renewcommand{\arraystretch}{1.2}
  \caption{ Results for $\Delta\epsK$. They are obtained using the
    RBC-UKQCD estimate for $\xi_\text{LD}$, the indirect method for
    $\xi_0$, the traditional method for $\eta_i$ of $c-t$ unitarity,
    the FLAG results for $\BK$, the AOF results for the Wolfenstein
    parameters, and so on.  }
  \label{tab:DepsK}
  \begin{ruledtabular}
    \begin{tabular}{l l l}
      year & Inclusive $\Vcb$ & Exclusive $\Vcb$ \\ \hline
      2015 & $0.33\sigma$     & $3.4\sigma$ \\
      2018 & $1.1\sigma$      & $4.2\sigma$ \\
      2025 & $1.4\sigma$      & $5.1\sigma$
    \end{tabular}
  \end{ruledtabular}
\end{table}